\documentclass[preprint,1p]{elsarticle}

\usepackage{amsmath}
\usepackage{amssymb}
\usepackage{enumerate}
\usepackage{centernot}
\usepackage{url}

\def\ci{\perp\!\!\!\perp}

\newtheorem{theorem}{Theorem}
\newtheorem{proposition}{Proposition}
\newtheorem{lemma}{Lemma}
\newtheorem{corollary}{Corollary}
\newdefinition{remark}{Remark}
\newcommand*{\QEDE}{\hfill\ensuremath{\square}}

\begin{document}
\begin{frontmatter}

%---- Title and abstract
%-----------------------------------------------------------------------------------
\title{Ridge Estimation of Inverse Covariance Matrices from High-Dimensional Data}

\author[EB,Math]{Wessel N. van Wieringen \corref{cor1}}
\ead{w.vanwieringen@vumc.nl}
\author[EB]{Carel F.W. Peeters} %\corref{cor2}}
\ead{cf.peeters@vumc.nl}

\cortext[cor1]{Principal corresponding author}
%\cortext[cor2]{Corresponding author}

\address[EB]{Department of Epidemiology \& Biostatistics,
       VU University medical center Amsterdam,
       Postbus 7057, 1007 MB Amsterdam, The Netherlands}
\address[Math]{Department of Mathematics,
       VU University Amsterdam,
       1081 HV Amsterdam, The Netherlands}

\begin{abstract}%
We study ridge estimation of the precision matrix in the
high-dimensional setting where the number of variables is large
relative to the sample size. We first review two archetypal ridge
estimators and note that their penalties do not coincide
with common quadratic ridge penalties. Subsequently, starting from a proper
$\ell_2$-penalty, analytic expressions are derived for two alternative
ridge estimators of the precision matrix. The alternative estimators
are compared to the archetypes with regard to eigenvalue shrinkage
and risk. The alternatives are also compared to the graphical lasso
within the context of graphical modeling. The comparisons may give
reason to prefer the proposed alternative estimators.
\end{abstract}

\begin{keyword}
graphical modeling \sep  high-dimensional precision matrix estimation \sep
multivariate normal \sep $\ell_2$-penalization \sep precision matrix
\end{keyword}

\end{frontmatter}

%---- Main text
%-----------------------------------------------------------------------------------
\section{Introduction}\label{Intro}
Let $\mathbf{Y}_{i}$, $i=1,\ldots,n$, be a $p$-dimensional random
variate drawn from $\mathcal{N}_{p}(\mathbf{0}, \mathbf{\Sigma})$.
The maximum likelihood (ML) estimator of the precision matrix
$\mathbf{\Omega} = \mathbf{\Sigma}^{-1}$ maximizes:
    \begin{eqnarray} \label{form.loglik}
    \mathcal{L}(\mathbf{\Omega}; \mathbf{S})  \propto \ln|\mathbf{\Omega}| - \mbox{tr}( \mathbf{S\Omega}),
    \end{eqnarray}
where $\mathbf{S}$ is the sample covariance estimate. If $n > p$,
the log-likelihood achieves its maximum for
$\hat{\mathbf{\Omega}}^{\mathrm{ML}}  = \mathbf{S}^{-1}$.

In the high-dimensional setting where $p > n$, the sample covariance
matrix is singular and its inverse is undefined. Consequently, so is
$\hat{\mathbf{\Omega}}^{\mathrm{ML}}$. A common workaround is the
addition of a penalty to the log-likelihood (\ref{form.loglik}). The
$\ell_1$-penalized estimation of the precision matrix was considered
almost simultaneously by \cite{YL07}, \cite{Bane2008},
\cite{Frie2008}, and \cite{Yuan08}. This (graphical) lasso estimate
of $\mathbf{\Omega}$ has attracted much attention due to the
resulting sparse solution. Juxtaposed to situations in which
sparsity is an asset are situations in which one is intrinsically
interested in more accurate representations of the high-dimensional
precision matrix. In addition, the true (graphical) model need not
be (extremely) sparse in terms of containing many zero elements. In
these cases we may prefer usage of a regularization method that
shrinks the estimated elements of the precision matrix
proportionally \citep{Fu1998} in possible conjunction with some form
of post-hoc element selection. It is such estimators we consider.

We thus study ridge estimation of the precision matrix. We first
review two archetypal ridge estimators and note that their
penalties do not coincide with what is perceived to be the common
ridge penalty (Section \ref{Archetype}). Subsequently, starting from
a common ridge penalty, analytic expressions are derived for
alternative ridge estimators of the precision matrix in Section
\ref{AltEst}. This section, in addition, studies properties of the
alternative estimators and proposes a method for choosing the
penalty parameter. In Section \ref{Compare} the alternative
estimators are compared to their corresponding archetypes w.r.t.
eigenvalue shrinkage. In addition, the risks of the various
estimators are assessed under multiple loss functions, revealing the
superiority of the proposed alternatives. Section \ref{Illustrate}
compares the alternative estimators to the graphical lasso in a
graphical modeling setting using oncogenomics data. This comparison
points to certain favorable behaviors of the proposed alternatives
with respect to loss, sensitivity, and specificity. In addition,
Section \ref{Illustrate} demonstrates that the alternative ridge
estimators yield more stable networks vis-\`{a}-vis the graphical
lasso, in particular for more extreme $p/n$ ratios. This section thus
provides empirical evidence in the graphical modeling setting of
what is tacitly known from regression
(subset selection) problems: ridge penalties coupled with post-hoc
selection may outperform the lasso. We conclude with
a discussion (Section \ref{Discuss}).
%%%%%%%%%%
%%%%%%%%%%

%%%%%%%%%%
%%%%%%%%%%
\section{Archetypal Ridge Estimators}\label{Archetype}
Ridge estimators of the precision matrix currently in use can be
roughly divided into two archetypes \citep[cf.][]{Ledo2004,SS05}. The first
archetypal form of ridge estimator commonly is a convex combination
of $\mathbf{S}$ and a positive definite (p.d.) target matrix
$\mathbf{\Gamma}$:
$\hat{\mathbf{\Omega}}^{\mathrm{I}}(\lambda_{\mathrm{I}}) =
[(1-\lambda_{\mathrm{I}}) \mathbf{S} + \lambda_{\mathrm{I}}
\mathbf{\Gamma}]^{-1}$, with $\lambda_{\mathrm{I}} \in (0,1]$. A common
(low-dimensional) target choice is $\mathbf{\Gamma}$ diagonal with
$(\mathbf{\Gamma})_{jj} = (\mathbf{S})_{jj}$ for $j=1, \ldots, p$. This
estimator has the desirable property of shrinking to
$\mathbf{\Gamma}^{-1}$ when $\lambda_{\mathrm{I}}=1$ (maximum
penalization). The estimator can be motivated from the bias-variance
tradeoff as it seeks to balance the high-variance, low-bias matrix
$\mathbf{S}$ with the lower-variance, higher-bias matrix
$\mathbf{\Gamma}$. It can also be viewed as resulting from the
maximization of the following penalized log-likelihood:
\begin{eqnarray} \label{form.penloglik_ridgeI}
\ln|\mathbf{\Omega}| - (1-\lambda_{\mathrm{I}}) \mbox{tr}( \mathbf{S\Omega}) - \lambda_{\mathrm{I}} \mbox{tr}( \mathbf{\Omega} \mathbf{\Gamma}).
\end{eqnarray}
The penalized log-likelihood (\ref{form.penloglik_ridgeI}) is
obtained from the original log-likelihood (\ref{form.loglik}) by the
replacement of $\mathbf{S}$ by $(1-\lambda_{\mathrm{I}}) \mathbf{S}$
and the addition of a penalty. The estimate
$\hat{\mathbf{\Omega}}^{\mathrm{I}}(\lambda_{\mathrm{I}})$ can thus
be viewed as a penalized ML estimate.

The second archetype finds its historical base in ridge regression,
a technique that started as an ad-hoc modification for dealing with
singularity in the least squares normal equations. The archetypal
second form of the ridge precision matrix estimate would be
$\hat{\mathbf{\Omega}}^{\mathrm{II}}(\lambda_{\mathrm{II}}) =
(\mathbf{S} + \lambda_{\mathrm{II}} \mathbf{I}_{p})^{-1}$
with $\lambda_{\mathrm{II}} \in (0, \infty)$. It can be motivated as
an ad-hoc fix of the singularity of $\mathbf{S}$ in the
high-dimensional setting, much like how ridge regression was
originally introduced by \cite{Hoer1970}. Alternatively, this
archetype too can be viewed as a penalized estimate, as it
maximizes \citep[see also][]{Wart08}:
\begin{eqnarray} \label{form.penloglik_ridgeII}
\ln|\mathbf{\Omega}| - \mbox{tr}( \mathbf{S\Omega}) - \lambda_{\mathrm{II}} \mbox{tr}( \mathbf{\Omega} \mathbf{I}_{p}).
\end{eqnarray}

The penalties in (\ref{form.penloglik_ridgeI}) and
(\ref{form.penloglik_ridgeII}) are non-concave (their second order
derivatives equal the null-matrix $\mathbf{0}$). This, however, poses
no problem under the restriction of a p.d. solution
$\mathbf{\Omega}$ as the Hessian of both
(\ref{form.penloglik_ridgeI}) and (\ref{form.penloglik_ridgeII})
equals $-\mathbf{\Omega}^{-2}$. More surprising is that neither
penalty of the two current archetypes resembles the
precision-analogy of what is commonly perceived as the ridge
$\ell_2$-penalty: $\frac{1}{2} \lambda \| \mathbf{\Omega} \|_2^2 =
\frac{1}{2} \lambda \sum_{j_1=1}^p \sum_{j_2=1}^p [ (\mathbf{\Omega})_{j_1,
j_2} ]^2$.

\begin{figure}[t]
\centering
\includegraphics[scale=.37]{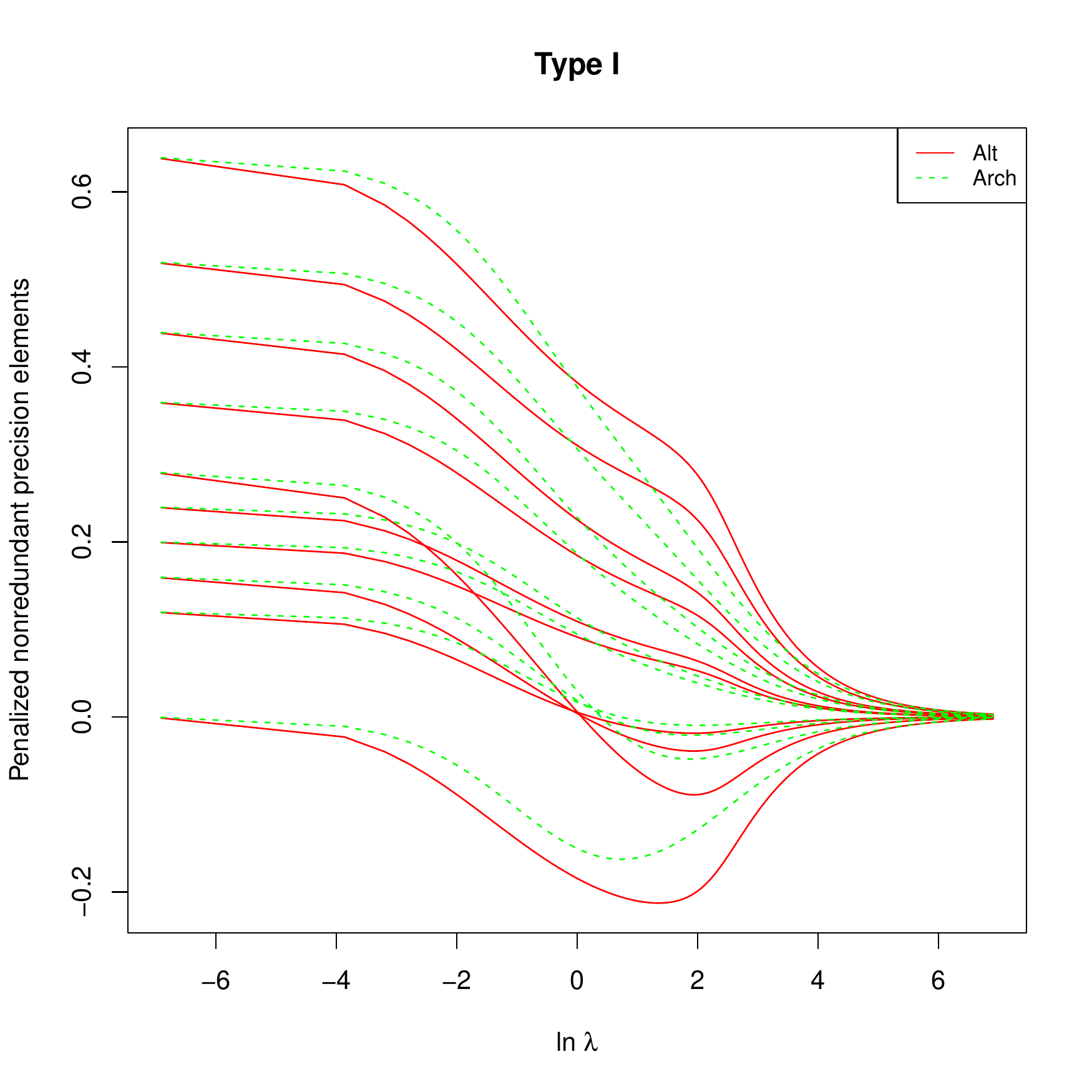}
\includegraphics[scale=.37]{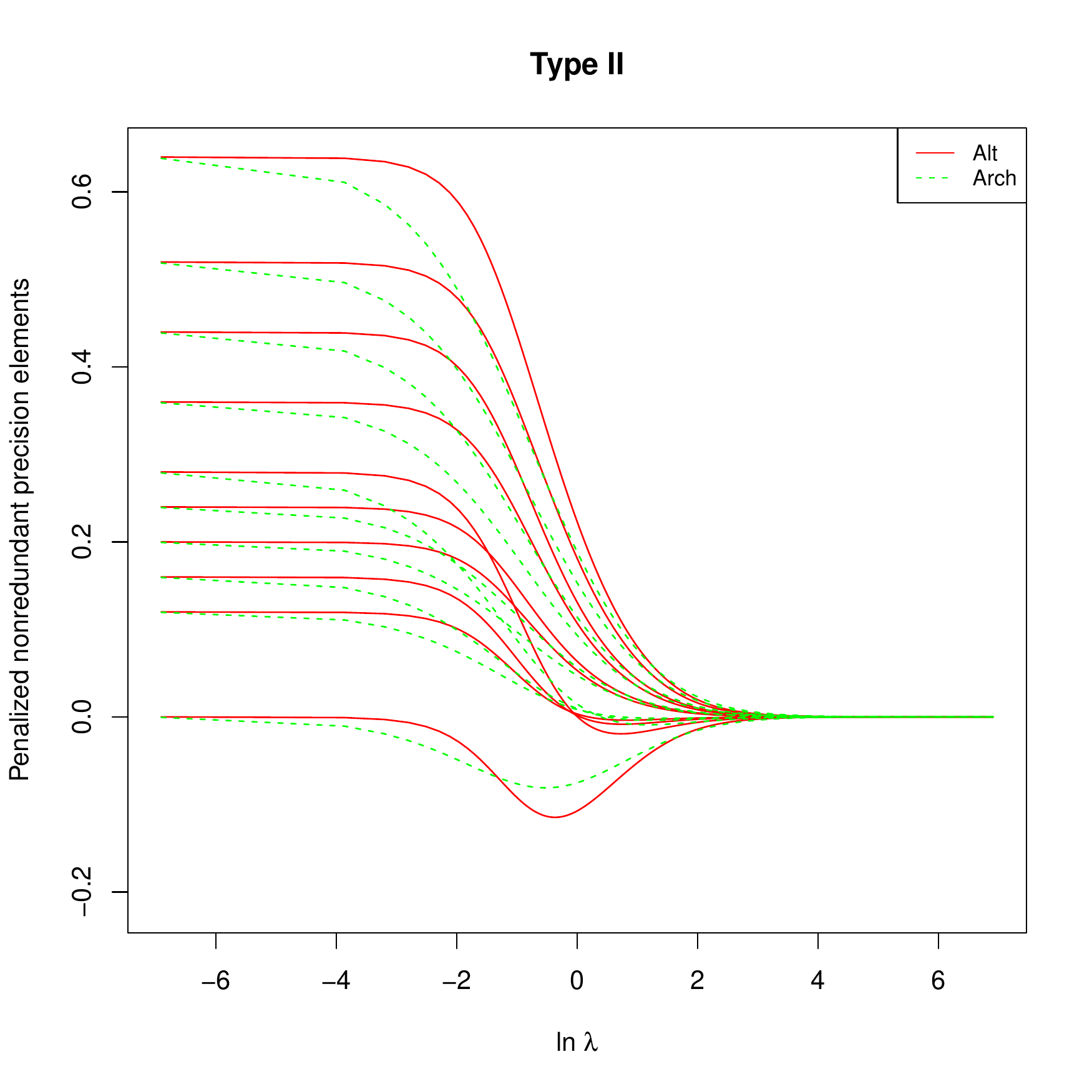}
\caption{Ridge coefficient paths of nonredundant off-diagonal
elements for the archetypal (dashed green) and alternative (solid red)
Type I (left panel) and Type II (right panel) ridge
estimators. The $5\times 5$ matrix $\mathbf{S}$ was generated as
$(\mathbf{S}^{-1})_{j_1,j_2} = [(j_1 \times j_2+1)
~\mbox{mod} ~21] / 25$ if $j_1 \neq j_2$ and
$(\mathbf{S}^{-1})_{j_1,j_2} = 1$ if $j_1 = j_2$. The target matrix
in the Type I case was taken to be the identity matrix
$\mathbf{I}_{5}$. The penalty parameter is generically indicated by
$\lambda$. For archetype-to-alternative scaling of the penalty
parameters under Type I and Type II estimation see Section
\ref{ShrinkMyEigenValue}.}\label{RidgeFig}
\end{figure}

The graphical lasso uses a penalty that is in line with the
$\ell_1$-penalty of lasso regression. It is a similar objective we
have in the remainder. We embark on the derivation of alternative
Type I and Type II (graphical) ridge estimators using a proper
$\ell_2$-penalty. Consider Figure \ref{RidgeFig} to get a flavor of
the behavior of both the archetypal ridge precision matrix estimators and
our alternatives (receiving analytic justification in Section
\ref{AltEst}). It is seen that ridge estimation based on a proper
ridge penalty induces (slight) differences in behavior. Differences
that will be shown to point to the preferability of the alternative
estimators in Section \ref{Compare}.

%%%%%%%%%%
%%%%%%%%%%

%%%%%%%%%%
%%%%%%%%%%
\section{Alternative Ridge Estimators of the Precision Matrix}\label{AltEst}
In this section we derive analytic expressions for alternative Type I and Type II
ridge precision estimators. In addition, we explore their moments (Section \ref{Moments})
and consistency (Section \ref{Consistency}) as well as methods for choosing the penalty
parameter (Section \ref{PenaltyChoice}). Proofs (as indeed all proofs in the remainder)
are deferred to \ref{Proofs}.

\subsection{Type I}\label{TypeI}
In this section an analytic expression for an alternative Type I
ridge precision estimator is given. Before arriving at a
proposition containing some properties of this estimator, we employ the following lemma:

\begin{lemma}[Alternative Type I ridge precision estimator]\label{GenRidgeAltLemma}
Amend the log-likelihood (\ref{form.loglik}) with the $\ell_2$-penalty
\begin{equation}\label{RidgePenal}
\frac{\lambda_a}{2} \mbox{\emph{tr}}\left[(\mathbf{\Omega} -
\mathbf{T})^{\mathrm{T}}(\mathbf{\Omega} - \mathbf{T})\right],
\end{equation}
with $\mathbf{T}$ denoting a symmetric p.d. target matrix, and
where $\lambda_{a} \in (0,\infty)$ denotes a penalty parameter.
Under given penalty, an alternative (penalized ML) Type I ridge
estimator is obtained as:
\begin{equation}\label{RidgeAltI}
\hat{\mathbf{\Omega}}^{\mathrm{I}a}(\lambda_{a}) =
\left\{\left[\lambda_{a}\mathbf{I}_{p} + \frac{1}{4}(\mathbf{S} -
\lambda_{a}\mathbf{T})^{2}\right]^{1/2} + \frac{1}{2}(\mathbf{S} -
\lambda_{a}\mathbf{T})\right\}^{-1}.
\end{equation}
\end{lemma}

\begin{proposition}\label{RidgeAltIProp}
Consider the alternative Type I ridge estimator (\ref{RidgeAltI})
from Lemma \ref{GenRidgeAltLemma}. For this estimator, the following properties hold:
\begin{enumerate}[i.]
  \item $\hat{\mathbf{\Omega}}^{\mathrm{I}a}(\lambda_{a}) \succ 0, ~\mbox{for all} ~\lambda_{a} \in (0, \infty)$;\vspace{-.2cm}
  \item $\lim_{\lambda_{a}\rightarrow 0^{+}} \hat{\mathbf{\Omega}}^{\mathrm{I}a}(\lambda_{a}) = \mathbf{S}^{-1}$;\vspace{-.2cm}
  \item $\lim_{\lambda_{a}\rightarrow \infty^{-}} \hat{\mathbf{\Omega}}^{\mathrm{I}a}(\lambda_{a}) = \mathbf{T}$.
\end{enumerate}
\end{proposition}

%Analogous to the archetypal I estimator, the right and left-hand
limits of the proposed estimator are the (possibly nonexistent)
inverse of the ML estimator $\mathbf{S}$ and a target matrix,
respectively. For a fuller understanding of the estimator
(\ref{RidgeAltI}), consider the following remarks.

\begin{remark}\label{AltITargetRemark}
The target matrix $\mathbf{T}$ from Lemma \ref{GenRidgeAltLemma} may
in principle be nonnegative definite (n.d.) for the statement to
hold. As should be clear from Proposition \ref{RidgeAltIProp},
however, choosing an n.d. target may lead to ill-conditioned
estimates in the limit. Moreover, from a shrinkage perspective, the
interpretability of a p.d. target may be deemed superior. Hence,
Lemma \ref{GenRidgeAltLemma} assumes the target matrix to be p.d.
(as does the archetypal Type I estimator). Section \ref{TypeII}
considers as a special case the n.d. choice $\mathbf{T} =
\mathbf{0}$, in order to arrive at an alternative for the archetypal
Type II estimator.
\end{remark}

\begin{remark}\label{AltIPenalRemark}
It may be noticed that the penalty term (\ref{RidgePenal})
amounts to a proper ridge penalty as
$\frac{\lambda_{a}}{2}\mbox{\emph{tr}}\left[(\mathbf{\Omega} -
\mathbf{T})^{\mathrm{T}}(\mathbf{\Omega} -
\mathbf{T})\right] = \frac{\lambda_{a}}{2}\|\mathbf{\Omega} - \mathbf{T}\|_{2}^{2}$.
When $\mathbf{T} = \mathbf{0}$, we obtain
$\frac{\lambda_{a}}{2}\|\mathbf{\Omega}\|_{2}^{2}$; a
special case that will be considered in Section \ref{TypeII}.
\end{remark}

\begin{remark}\label{AltICondRemark}
From Proposition \ref{RidgeAltIProp} it is clear that
(\ref{RidgeAltI}) is always p.d. when $\lambda_{a} \in (0,
\infty)$. However, as with any regularized covariance or precision
estimator, the estimate is not necessarily \emph{well-conditioned}
(in terms of, say, the spectral condition number) for any
$\lambda_{a} \in (0, \infty)$ when $\mathbf{S}$ is
ill-behaved. To obtain a well-conditioned estimate in such
situations, one should choose $\lambda_{a}$ not too close
to zero. In order to choose an optimal value of
$\lambda_{a}$ for a problem at hand, one can employ
(approximate) cross-validation or information criteria
(see Section \ref{PenaltyChoice}).
\end{remark}

\begin{remark}\label{AltICovEstRemark}
\begin{sloppypar}
Lemma \ref{GenRidgeAltLemma} considers regularized estimation of
the precision matrix. It may also provide an alternative Type I
regularized estimator for the covariance matrix, by entertaining
\[
[\hat{\mathbf{\Omega}}^{\mathrm{I}a}(\lambda_{a})]^{-1}
\equiv \hat{\mathbf{\Sigma}}^{\mathrm{I}a}(\lambda_{a}) =
\left[\lambda_{a}\mathbf{I}_{p} + \frac{1}{4}(\mathbf{S} -
\lambda_{a}\mathbf{T})^{2}\right]^{1/2} +
\frac{1}{2}(\mathbf{S} - \lambda_{a}\mathbf{T}).\nonumber
\]
Then: (i)
$\hat{\mathbf{\Sigma}}^{\mathrm{I}a}(\lambda_{a}) \succ 0,
~\mbox{for all} ~\lambda_{a} > 0$; (ii)
$\lim_{\lambda_{a}\rightarrow 0^{+}}
\hat{\mathbf{\Sigma}}^{\mathrm{I}a}(\lambda_{a}) =
\mathbf{S}$; (iii) $\lim_{\lambda_{a}\rightarrow
\infty^{-}}
\hat{\mathbf{\Sigma}}^{\mathrm{I}a}(\lambda_{a}) =
\mathbf{T}^{-1}$. Say one wishes to shrink to a p.d. covariance
target $\mathbf{C}$, one only has to specify $\mathbf{T} =
\mathbf{C}^{-1}$ in this case.
\end{sloppypar}
\end{remark}

\begin{remark}\label{NoInversionRemark}
We note that (\ref{RidgeAltI}) can also be obtained without inversion, by noticing
\begin{eqnarray}\nonumber
\hat{\mathbf{\Omega}}^{\mathrm{I}a}(\lambda_{a}) =
\frac{1}{\lambda_{a}}\left[\hat{\mathbf{\Sigma}}^{\mathrm{I}a}(\lambda_{a}) - (\mathbf{S} - \lambda_{a}\mathbf{T})\right].
\end{eqnarray}
The basis for this claim is expression (\ref{PCidentity}) from Section \ref{Moments}.
\end{remark}

\subsection{Type II}\label{TypeII}
An alternative Type II ridge estimator for the precision matrix can
be found as a special case of Lemma \ref{GenRidgeAltLemma}:

\begin{corollary}[Alternative Type II ridge precision estimator]\label{RidgeAltIIColl}
Consider the alternative Type I ridge estimator (\ref{RidgeAltI})
from Lemma \ref{GenRidgeAltLemma}. An alternative ridge proper Type
II estimator is obtained by choosing $\mathbf{T} = \mathbf{0}$, such that
\begin{equation}\label{RidgeAltII}
\hat{\mathbf{\Omega}}^{\mathrm{II}a}(\lambda_{a}) =
\left\{\left[\lambda_{a}\mathbf{I}_{p} +
\frac{1}{4}\mathbf{S}^{2}\right]^{1/2} + \frac{1}{2}\mathbf{S}\right\}^{-1}.
\end{equation}
For this estimator, the following properties hold:
\begin{enumerate}[i.]
  \item $\hat{\mathbf{\Omega}}^{\mathrm{II}a}(\lambda_{a}) \succ 0, ~\mbox{for all} ~\lambda_{a} \in (0, \infty)$;\vspace{-.2cm}
  \item $\lim_{\lambda_{a}\rightarrow 0^{+}} \hat{\mathbf{\Omega}}^{\mathrm{II}a}(\lambda_{a}) = \mathbf{S}^{-1}$;\vspace{-.2cm}
  \item $\lim_{\lambda_{a}\rightarrow \infty^{-}} \hat{\mathbf{\Omega}}^{\mathrm{II}a}(\lambda_{a}) = \mathbf{0}$.
\end{enumerate}
\end{corollary}

Similar to the archetypal II estimator, the right and left-hand
limits are the (possibly nonexistent) inverse of the ML estimator
$\mathbf{S}$ and the null-matrix, respectively. The alternative Type
II analogies of Remarks
\ref{AltICondRemark}--\ref{NoInversionRemark} hold for
(\ref{RidgeAltII}). Note that the estimator (\ref{RidgeAltII}) was
also considered by \cite{Wit09} in a different setting.

\subsection{Moments}\label{Moments}
The explicit expressions for the alternative (Type I and II) ridge estimators facilitate the study of their properties. For instance, the moments of the ridge covariance and precision estimators can -- in principle -- be evaluated numerically to any desired degree of accuracy. Consider the following exemplification. With respect to the alternative Type I estimator we write:
\begin{eqnarray}\nonumber\label{RW}
\hat{\mathbf{\Sigma}}^{\mathrm{I}a}(\lambda_{a}) = \sqrt{\lambda_{a}}\left[ (\mathbf{I}_{p} + \mathbf{U}^2)^{1/2} + \mathbf{U}\right] ,
\end{eqnarray}
where $\mathbf{U} =  ( \mathbf{S} - \lambda_{a} \mathbf{T}) \, / \, (2 \sqrt{\lambda_{a}})$. Express the $(1 + x^2)^{1/2}$ term as a binomial series to obtain the series representation of the ridge covariance estimator:
\begin{eqnarray}\nonumber\label{BSeries}
\hat{\mathbf{\Sigma}}^{\mathrm{I}a}(\lambda_{a}) = \sqrt{\lambda_{a}} \mathbf{U} + \sqrt{\lambda_{a}} \sum_{q=0}^{\infty} { 1/2 \choose q } \mathbf{U}^{2q}.
\end{eqnarray}
Now, taking the expectation of the right-hand side yields the first moment of the alternative Type I ridge covariance estimator. To evaluate this expectation note that (under normality) $\mathbf{S}$ follows a (singular) Wishart distribution, assume $\mathbf{T}$ to be non-random, and restrict the binomial series to the degree that produces the desired accuracy. It then suffices to plug in the required moments of the Wishart distribution.

From the moments of the ridge covariance estimator one can directly obtain the moments of the ridge precision estimator. Hereto we need the identity:
\begin{eqnarray}\label{identity}
2 \sqrt{\lambda_{a}} \mathbf{U} = \sqrt{\lambda_{a}} \left[ (\mathbf{I}_{p} + \mathbf{U}^2)^{1/2} + \mathbf{U}\right] - \sqrt{\lambda_{a}} \left[ (\mathbf{I}_{p} + \mathbf{U}^2)^{1/2} + \mathbf{U}\right]^{-1},
\end{eqnarray}
with $\mathbf{U}$ as above. This equality is immediate after noting that all terms have the same eigenvectors and using ready algebra to prove the identity $2 x = x + (1+x^2)^{1/2}  - [x + (1+x^2)^{1/2}]^{-1}$, which applies to each eigenvalue in the eigen-decomposition (see also Section \ref{ShrinkMyEigenValue}) of (\ref{identity}) separately. Reformulated we then have:
\begin{eqnarray}\label{PCidentity}
\mathbf{S} - \lambda_{a} \mathbf{T} = \hat{\mathbf{\Sigma}}^{\mathrm{I}a}(\lambda_{a}) - \lambda_{a} \hat{\mathbf{\Omega}}^{\mathrm{I}a}(\lambda_{a}).
\end{eqnarray}
This identity thus yields, via the moments of the alternative Type I
ridge covariance matrix, the moments of the alternative Type I ridge
precision matrix. The moments of the alternative Type II estimator
can be obtained when considering $\mathbf{T}$ to be the null-matrix.

Being able to evaluate the moments facilitates, e.g., the
approximation of the bias of the proposed ridge estimators. Hereto
assume $\mathbf{Y}_i \sim \mathcal{N}_{p}(\mathbf{0},
\mathbf{\Sigma})$ for $i = 1, \ldots n$. Define the sample
covariance matrix $\mathbf{S} = \frac{1}{n} \sum_{i=1}^n
\mathbf{Y}_i \mathbf{Y}_i^{\mathrm{T}}$. Then, it is well-known that
$n \mathbf{S}$ follows the Wishart distribution
$\mathcal{W}_{p}(\mathbf{\Sigma}, n)$. Recently, \cite{Wish} have
shown how $\mathbb{E}(n^b \mathbf{S}^b)$ may be derived analytically
when $b \in \mathbb{Z}$. Their results are exploited here to
approximate the bias of the proposed ridge estimators. When we
ignore terms of order three and higher and limit ourselves to the
type II estimator with $\mathbf{T} = \mathbf{0}$, the expectation
may be approximated by (see also Section 1 of the Supplementary
Material):
\begin{eqnarray*}
\mathbb{E}\left[ \hat{\mathbf{\Sigma}}^{\mathrm{II}a}(\lambda_{a})
\right] & \approx & \frac{1}{2} \mathbb{E}(\mathbf{S}) +
\sqrt{\lambda_{a}} \mathbf{I}_{p} + \frac{1}{8 \sqrt{\lambda_{a}}}
\mathbb{E}( \mathbf{S}^{2})
\\
& = &  \frac{1}{2} \mathbf{\Sigma} + \sqrt{\lambda_{a}}
\mathbf{I}_{p}  + \frac{1}{8 \sqrt{\lambda_a}} \left[ \frac{n+1}{n}
\mathbf{\Sigma}^2 + \frac{1}{n} \mbox{tr} ( \mathbf{\Sigma})
\mathbf{\Sigma}\right],
\end{eqnarray*}
in which the expectations of $\mathbf{S}$ and $\mathbf{S}^2$ are
obtained from \cite{Wish}. Section 1 of the Supplementary Material
contains a higher-order approximation and a simulation illustrating
the accuracy of the approximation.

\subsection{Consistency}\label{Consistency}
We will show that the alternative Type I  ridge estimator
(\ref{RidgeAltI}) is consistent under fixed-dimension asymptotics.
To make this explicit, we (temporarily) modify the notation. Let
$\mathbf{S}_n$ be the sample covariance matrix with index $n$
indicating the sample size. Furthermore, the penalty parameter is
now denoted $\lambda_{a,n}$. This explicates the fact that the
penalty parameter is chosen in a data-driven fashion (cf. Section
\ref{PenaltyChoice}) and thus depends on the sample size. In
particular, it will be assumed that $\lambda_{a,n}$ converges (in
some sense) to zero as $n \rightarrow \infty^{-}$. This reflects the
decreasing necessity to regularize the (inverse) covariance
estimator as the sample size increases. Finally, let
$\hat{\mathbf{\Sigma}}_n^{\mathrm{I}a} (\lambda_{a,n})$ be the
alternative ridge covariance estimator (see Remark
\ref{AltICovEstRemark}) with $\mathbf{S}$ and $\lambda_a$ replaced
by $\mathbf{S}_n$ and $\lambda_{a,n}$.

In showing consistency, we need theasymptotic unbiasedness of our
estimator. This property is warranted by the following lemma:

\begin{lemma}[Asymptotic unbiasedness] \label{prop.asympBias}
Let $\mathbf{S}_n$ be the sample covariance matrix from a sample
$\mathbf{Y}_1, \ldots, \mathbf{Y}_n$ drawn from
$\mathcal{N}_{p}(\mathbf{0}, \mathbf{\Sigma})$. Denote by
$\lambda_{a,n}$ a nonnegative random variable that converges almost
surely to zero and by $\mathbf{T}$ a nonrandom p.d. symmetric
matrix. Then:
\begin{eqnarray*}
\lim_{n \rightarrow \infty^{-} }  \mathbb{E} \left[ \hat{\mathbf{\Sigma}}_n^{\mathrm{I}a} (\lambda_{a,n})\right]  \longrightarrow \lim_{n \rightarrow \infty^{-} } \mathbb{E}( \mathbf{S}_n) =  \mathbf{\Sigma}.
\end{eqnarray*}
Simultaneously, the expectation of its inverse
$\hat{\mathbf{\Omega}}_n^{\mathrm{I}a} (\lambda_{a,n})$ tends to
$\mathbf{\Omega} = \mathbf{\Sigma}^{-1}$ as $n \rightarrow
\infty^{-}$.
\end{lemma}
\noindent Lemma \ref{prop.asympBias} follows directly from
application of the continuous mapping theorem, the Portmanteau lemma
and Slutsky's lemma \citep[see, e.g., Theorem 2.3, Lemma 2.2, and
Lemma 2.8 in][]{Vaart98}. By virtue of the same asymptotic results,
the lemma may be generalized to allow $\mathbf{T}$ to depend on
data, as long as the data-dependent target $\mathbf{T}_n$ converges
(almost surely) to some $\mathbf{T}$.

Lemma \ref{prop.asympBias} is conducive in proving the consistency
result (note that asymptotic unbiasedness and consistency of the
alternative Type II estimator (\ref{RidgeAltII}) follow as special
cases of Lemma \ref{prop.asympBias} and Proposition
\ref{prop.consistency}):

\begin{proposition}[Consistency]\label{prop.consistency}
Let $\mathbf{S}_n$ be the sample covariance matrix from a sample
$\mathbf{Y}_1, \ldots, \mathbf{Y}_n$ drawn from
$\mathcal{N}_{p}(\mathbf{0}, \mathbf{\Sigma})$. Denote by
$\lambda_{a,n}$ a nonnegative random variable that converges almost
surely to zero and by $\mathbf{T}$ a nonrandom p.d. symmetric
matrix. Then:
\begin{eqnarray*}
\lim_{n \rightarrow \infty^{-}} \mathbb{E} \left( \| \hat{\mathbf{\Sigma}}_n^{\mathrm{I}a} (\lambda_{a,n}) - \mathbf{\Sigma} \|_F^2 \right) = 0,
\end{eqnarray*}
where $\| \cdot \|_F$ denotes the Frobenius norm. Simultaneously,
$\hat{\mathbf{\Omega}}_n^{\mathrm{I}a}(\lambda_{a,n})$ consistently
estimates $\mathbf{\Omega} = \mathbf{\Sigma}^{-1}$.
\end{proposition}

The consistency result in Proposition \ref{prop.consistency} takes
$p$ to be fixed, and thus does not concern increasing-dimension
asymptotics (in which $p$ also tends to infinity). This is motivated
by practice. We have an applicatory focus on the reconstruction of
(molecular) interaction networks (see also Section
\ref{Illustrate}). The (maximum) number of variates of such systems
is fixed. As such, the consistency result above is deemed
appropriate.

\subsection{Choosing $\lambda_{a}$}\label{PenaltyChoice}
A well-informed choice of the penalty parameter $\lambda_{a}$ is
crucial in applications. The literature
contains many proposals for selecting an (in some sense) optimal
value for the penalty parameter in (precision) regularization
problems. These can be classified \citep[see][]{Abbruz2014} in methods
aiming at model selection consistency (e.g., BIC, EBIC), and methods
that aim to maximize predictive power (e.g., cross-validation, AIC).
As the $\ell_2$-penalty does not automatically induce sparsity in
the estimate, we are not after model selection consistency. Rather,
in our case it is natural to seek loss efficiency.

While both cross-validation (CV) and AIC \citep{Aik73} have similar
asymptotic properties in terms of minimizing Kullback-Leibler
divergence, the data-driven nature of the former makes it prone to
have superior behavior in terms of accuracy. The $K$-fold CV score for a generic regularized
estimate $\hat{\mathbf{\Omega}}(\lambda)$ based on the generic fixed penalty
$\lambda$ can be given as:
\[
\varphi^{K}(\lambda) = \sum_{k=1}^{K} n_k \left\{
-\ln|\hat{\mathbf{\Omega}}(\lambda)_{-k}| +
\mbox{tr}[\hat{\mathbf{\Omega}}(\lambda)_{-k}\mathbf{S}_k] \right\},
\]
where $n_k$ is the size of subset $k$, for $k = 1, \ldots, K$
disjoint subsets. Further, $\mathbf{S}_k$ denotes the sample
covariance matrix based on subset $k$, while
$\hat{\mathbf{\Omega}}(\lambda)_{-k}$ denotes the estimated regularized precision matrix
on all samples not in $k$. Highest predictive accuracy can be
obtained by choosing $n_k = 1$, such that $K = n$. This is known as
leave-one-out CV (LOOCV). Unfortunately, LOOCV (as $K$-fold CV in general)
is computationally demanding for large $p$ and/or large $n$.

Recently, \cite{Lian2011} and \cite{Vuja2014} derived, based on the
log-likelihood of the precision, an approximate solution to the
LOOCV score. Based on their work, the approximate LOOCV score for
fixed $\lambda_{a}$, $\tilde{\varphi}^{n}(\lambda_{a})$, is given
for the alternative Type I ridge estimator as:
\begin{equation}\label{ApproximateCV}
\tilde{\varphi}^{n}(\lambda_{a}) =
-\frac{1}{n}\mathcal{L}[\hat{\mathbf{\Omega}}^{\mathrm{I}a}(\lambda_{a});
\mathbf{S}] + \frac{1}{2n(n-1)}\sum_{i=1}^{n} \gamma_{i},
\end{equation}
with
\[
\gamma_{i} = \sum_{j_{1}=1}^{p}\sum_{j_{2}=1}^{p} \left\{ \left[
[\hat{\mathbf{\Omega}}^{\mathrm{I}a}(\lambda_{a})]^{-1} -
\mathbf{Y}_i\mathbf{Y}_i^{\mathrm{T}} \right] \circ \left[
\hat{\mathbf{\Omega}}^{\mathrm{I}a}(\lambda_{a}) \left( \mathbf{S} -
\mathbf{Y}_i\mathbf{Y}_i^{\mathrm{T}} \right)
\hat{\mathbf{\Omega}}^{\mathrm{I}a}(\lambda_{a}) \right]
\right\}_{j_{1},j_{2}},
\]
and where $\circ$ denotes the Hadamard product. Naturally, the
approximate LOOCV score for the alternative Type II ridge estimator
can be obtained by replacing
$\hat{\mathbf{\Omega}}^{\mathrm{I}a}(\lambda_{a})$ in
(\ref{ApproximateCV}) by
$\hat{\mathbf{\Omega}}^{\mathrm{II}a}(\lambda_{a})$. We propose to
choose $\lambda_{a}^*$ such that $\lambda_{a}^* =
\arg\min_{\lambda_{a} \in
\mathbb{R}^{+}}\tilde{\varphi}^{n}(\lambda_{a})$, which relates to
the minimization of Kullback-Leibler divergence and the maximization
of predictive accuracy. The expression $\tilde{\varphi}^{n}(\lambda)$ is
computationally efficient, requiring only a single matrix inversion
(as opposed to $n$ inversions for $\varphi^{n}(\lambda)$). In addition, the Hadamard
product has an efficient computational implementation \citep[see][]{Vuja2014}.

\begin{remark}
We note that only a single spectral decomposition and a single matrix inversion are required in order to obtain the complete solution path (over any $\lambda_a$ in the feasible domain) for the alternative Type II estimator and the alternative Type I estimator under a scalar matrix target choice (cf. Section \ref{ShrinkMyEigenValue}). This implies that, in these cases, the computation of $\tilde{\varphi}^{n}(\lambda_{a})$ over the (complete) solution path is particularly efficient. This efficiency, coupled with the benefits of knowing the full solution path, may be deemed to rival the benefits of a solution under an analytic choice of $\lambda_{a}$ (see also next remark).
\end{remark}

\begin{remark}
There exist analytic solutions to determining an
optimal value for the penalty parameter.
\cite{Ledo2003}, e.g., determine analytically, under a modified Frobenius loss,
the optimal value for the penalty parameter in an archetypal Type I
setting under certain choices of $\mathbf{T}$. For practical applications, however, one still needs to
approximate this optimal value, requiring variances and covariances
of the individual entries of $\mathbf{S}$ \citep{SS05}. When the
variable to observation ratio grows more extreme, the approximation
may propose (overly) conservative or even negative penalty values as
optimal \citep[cf.][]{DK01,SS05}, giving us reason to prefer the computationally
friendly, data-driven approach from above. In addition, (\ref{ApproximateCV}) is
generic, meaning it can be used under any p.d. choice of $\mathbf{T}$.
\end{remark}

\begin{sloppypar}
The estimates $\hat{\mathbf{\Omega}}^{\mathrm{I}a}(\lambda^{*}_{a})$ and $\hat{\mathbf{\Omega}}^{\mathrm{II}a}(\lambda^{*}_{a})$ may facilitate methods of (high-dimensional) data analysis in need of a precision (or covariance) matrix that is not (necessarily) sparse (cf. Sections \ref{Compare} and \ref{Discuss}). They may also be of interest in situations in which sparsity is required, such as graphical modeling. One may pair, in such situations, the proposed estimates with \emph{a posteriori} methods of support determination (cf. Section \ref{Illustrate}).
\end{sloppypar}
%%%%%%%%%%
%%%%%%%%%%

%%%%%%%%%%
%%%%%%%%%%
\section{Comparing Alternative and Archetypal Ridge Estimation}\label{Compare}
In this section the proposed alternative Type I and Type II ridge estimators are compared
to their corresponding archetypes w.r.t. eigenvalue shrinkage
(Section \ref{ShrinkMyEigenValue}). Moreover, the alternative and
archetypal estimators are subjected to a risk comparison (Section \ref{RC}).

\subsection{Eigenvalue Shrinkage}\label{ShrinkMyEigenValue}
The alternative Type I estimator (\ref{RidgeAltI}) is, as its
archetypal counterpart, rotation equivariant when choosing the target to be
a scalar matrix $\mathbf{T} = \psi\mathbf{I}_p$, with (for the alternative
Type I estimator) $\psi \in [0,\infty)$. That is, the effect of
the ridge penalty on the precision estimate is then equivalent to
shrinkage of the eigenvalues of the unpenalized estimate
$\mathbf{S}^{-1}$. To see this, let the eigen-decomposition of
$\mathbf{S}$ be $\mathbf{VDV}^{\mathrm{T}}$ where
$\mathbf{D}$ is a diagonal matrix with the eigenvalues of
$\mathbf{S}$ on the diagonal and $\mathbf{V}$ denotes the matrix
that contains the corresponding eigenvectors as columns. The
orthogonality of $\mathbf{V}$ implies $\mathbf{VV}^{\mathrm{T}} =
\mathbf{V}^{\mathrm{T}} \mathbf{V} = \mathbf{I}_{p}$. We then
rewrite, using $\mathbf{T} = \mathbf{I}_{p}$ for notational
convenience, the inverse of (\ref{RidgeAltI}) as follows:
\begin{align}\label{TypeIDecomp}\nonumber
[ \hat{\mathbf{\Omega}}^{\mathrm{I}a}(\lambda_{a}) ]^{-1}
& = \left[\lambda_{a}\mathbf{V}\mathbf{V}^{\mathrm{T}} +
\frac{1}{4}(\mathbf{VDV}^{\mathrm{T}} -
\lambda_{a}\mathbf{V}\mathbf{V}^{\mathrm{T}})^{2}\right]^{1/2}
+ \frac{1}{2}(\mathbf{VDV}^{\mathrm{T}} -
\lambda_{a}\mathbf{V}\mathbf{V}^{\mathrm{T}})
\\
& = \mathbf{V}\left\{\left[\lambda_{a}\mathbf{I}_{p} +
\frac{1}{4}(\mathbf{D} -
\lambda_{a}\mathbf{I}_{p})^{2}\right]^{1/2} +
\frac{1}{2}(\mathbf{D} -
\lambda_{a}\mathbf{I}_{p})\right\}\mathbf{V}^{\mathrm{T}},
\end{align}
making clear that the ridge penalty deals with singularity and
ill-conditioning through shrinkage of the eigenvalues of
$\mathbf{S}^{-1}$. The alternative Type II estimator (\ref{RidgeAltII}) also has the
property of being rotation equivariant. This can be seen by:
\begin{eqnarray}\label{TypeIIDecomp}
[ \hat{\mathbf{\Omega}}^{\mathrm{II}a}(\lambda_{a})
]^{-1}
%& = & \left( \lambda_{\mathrm{II}a} \mathbf{VV}^{\mathrm{T}}
%+ \frac{1}{4} \mathbf{VDV}^{\mathrm{T}}\mathbf{VDV}^{\mathrm{T}}
%\right)^{1/2} + \frac{1}{2} \mathbf{VDV}^{\mathrm{T}}
%\\
= \mathbf{V} \left[ \left( \lambda_{a}
\mathbf{I}_{p} + \frac{1}{4} \mathbf{D}^2\right)^{1/2}  +
\frac{1}{2} \mathbf{D} \right] \mathbf{V}^{\mathrm{T}}.
\end{eqnarray}
The equivariance property can be used in the
comparison of eigenvalue shrinkage between the archetypes and
alternatives. The following claims summarize:

\begin{proposition}\label{AltIEIGprop}
Let the regularization parameters of the archetypal and alternative
Type I ridge estimators -- $\lambda_{\mathrm{I}}$ and
$\lambda_{a}$ respectively -- map to the same scale. That
is, choose $\lambda_{\mathrm{I}} = 1 - 1/(\lambda_{a} +
1)$. In addition, consider a p.d. scalar matrix as the low-dimensional
target matrix $\mathbf{T}$ and let the archetypal Type I estimator
have the same target in the precision sense, i.e., $\mathbf{\Gamma}^{-1} = \mathbf{T}$.
Then the alternative estimator
$\hat{\mathbf{\Omega}}^{\mathrm{I}a}(\lambda_{a})$ displays
shrinkage of the eigenvalues of $\mathbf{S}^{-1}$ that is at least
as heavy as the shrinkage propagated by the archetypal estimator
$\hat{\mathbf{\Omega}}^{\mathrm{I}}(\lambda_{\mathrm{I}})$.
\end{proposition}

\begin{proposition}\label{AltIIEIGprop}
Let the regularization parameters of the archetypal and alternative
Type II ridge estimators -- $\lambda_{\mathrm{II}}$ and
$\lambda_{a}$ respectively -- map to the same scale. That
is, choose $\lambda_{a} = \lambda_{\mathrm{II}}^{2}$.
Then the archetypal estimator
$\hat{\mathbf{\Omega}}^{\mathrm{II}}(\lambda_{\mathrm{II}})$
displays shrinkage of the eigenvalues of $\mathbf{S}^{-1}$ that is
at least as heavy as the shrinkage propagated by the alternative
estimator
$\hat{\mathbf{\Omega}}^{\mathrm{II}a}(\lambda_{a})$.
\end{proposition}

\begin{corollary}\label{AltIILIKEcor}
The eigenvalue inequality of Proposition \ref{AltIIEIGprop} implies:
\begin{eqnarray*}
\mathcal{L}[\hat{\mathbf{\Omega}}^{\mathrm{II}}(\lambda_{\mathrm{II}});
\mathbf{S}] & \leq &
\mathcal{L}[\hat{\mathbf{\Omega}}^{\mathrm{II}a}
(\lambda_{a}); \mathbf{S}].
\end{eqnarray*}
\end{corollary}

The alternative Type I estimator displays \emph{faster} shrinkage to
the target $\mathbf{T}$ than the archetypal Type I estimator. The
alternative estimator then can be expected to have lower risk (in
terms of, say, quadratic loss) than its archetypal counterpart when
the (low-dimensional) target is an adequate representation of the
true precision matrix. In such cases it can be shown under mild assumptions that, analogous to
Corollary \ref{AltIILIKEcor}, $\mathcal{L}[\hat{\mathbf{\Omega}}^{\mathrm{I}}(\lambda_{\mathrm{I}});
\mathbf{S}] \leq  \mathcal{L}[\hat{\mathbf{\Omega}}^{\mathrm{I}a}(\lambda_{a}); \mathbf{S}]$.
In absence of a natural target $\mathbf{T}$,
Type II estimators are an option. It is seen from proposition
\ref{AltIIEIGprop} that, as opposed to the Type I situation, the
alternative Type II estimator displays \emph{slower} shrinkage to
the null-matrix than the archetypal Type II estimator. As the
limiting null-matrix can indeed never be a good representation of
the true precision matrix, the alternative Type II estimator can also be
expected to have lower risk than its archetypal counterpart. The behavior
of the alternative Type I and Type II estimators with regard to shrinkage
rate may initially seem contradictory when evaluating Propositions
\ref{AltIEIGprop} and \ref{AltIIEIGprop}. It is not if we notice that
the penalty parameter $\lambda_{a}$ is more influential in the Type
I alternative as its effect is not diluted by a null $\mathbf{T}$.
The topics of Loss and Risk are explored in the next subsection.

\subsection{Risk}\label{RC} The risks of the alternative
Type I and Type II estimators for the precision matrix are compared
to that the of Type I and II archetypes. Let $\mathbf{\Omega}$
denote a generic $(p \times p)$ population precision matrix and let
$\hat{\mathbf{\Omega}}(\lambda)$ denote a generic ridge estimator of
the precision matrix under generic regularization parameter
$\lambda$. The following loss functions are then considered in risk
evaluation:
\begin{description}
  \item[a.] Squared Frobenius loss, given by:
  \begin{equation}\nonumber
    L_{F}[\hat{\mathbf{\Omega}}(\lambda), \mathbf{\Omega}] = \|\hat{\mathbf{\Omega}}(\lambda) -
    \mathbf{\Omega}\|_{F}^{2};
    \end{equation}
  \item[b.] Quadratic loss, given by:
    \begin{equation}\nonumber
    L_{Q}[\hat{\mathbf{\Omega}}(\lambda), \mathbf{\Omega}] = \|\hat{\mathbf{\Omega}}(\lambda)
    \mathbf{\Omega}^{-1} - \mathbf{I}_{p}\|_{F}^{2}.
    \end{equation}
\end{description}
The risk $\mathcal{R}_{f}$ of the estimator
$\hat{\mathbf{\Omega}}(\lambda)$ given a loss function $L_{f}$, $f
\in \{F, Q\}$, is then defined as the expected loss:
\begin{equation}\nonumber
\mathcal{R}_{f}[\hat{\mathbf{\Omega}}(\lambda)] =
\mathbb{E}\{L_{f}[\hat{\mathbf{\Omega}}(\lambda),
\mathbf{\Omega}]\},
\end{equation}
which is approximated by the median of losses over repeated simulation runs.

\begin{figure}[h]
    \centering
    \includegraphics[scale=.37]{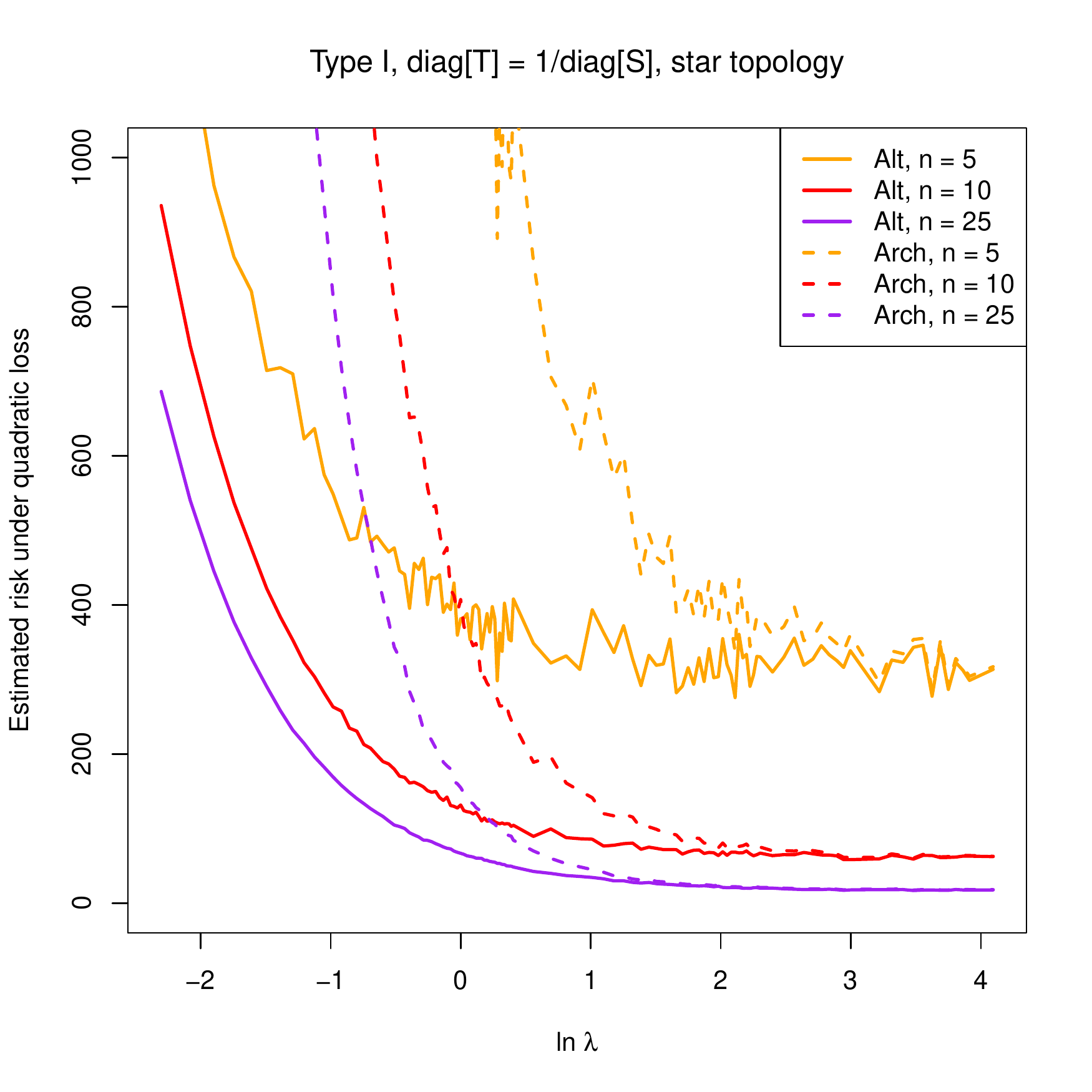}
    \includegraphics[scale=.37]{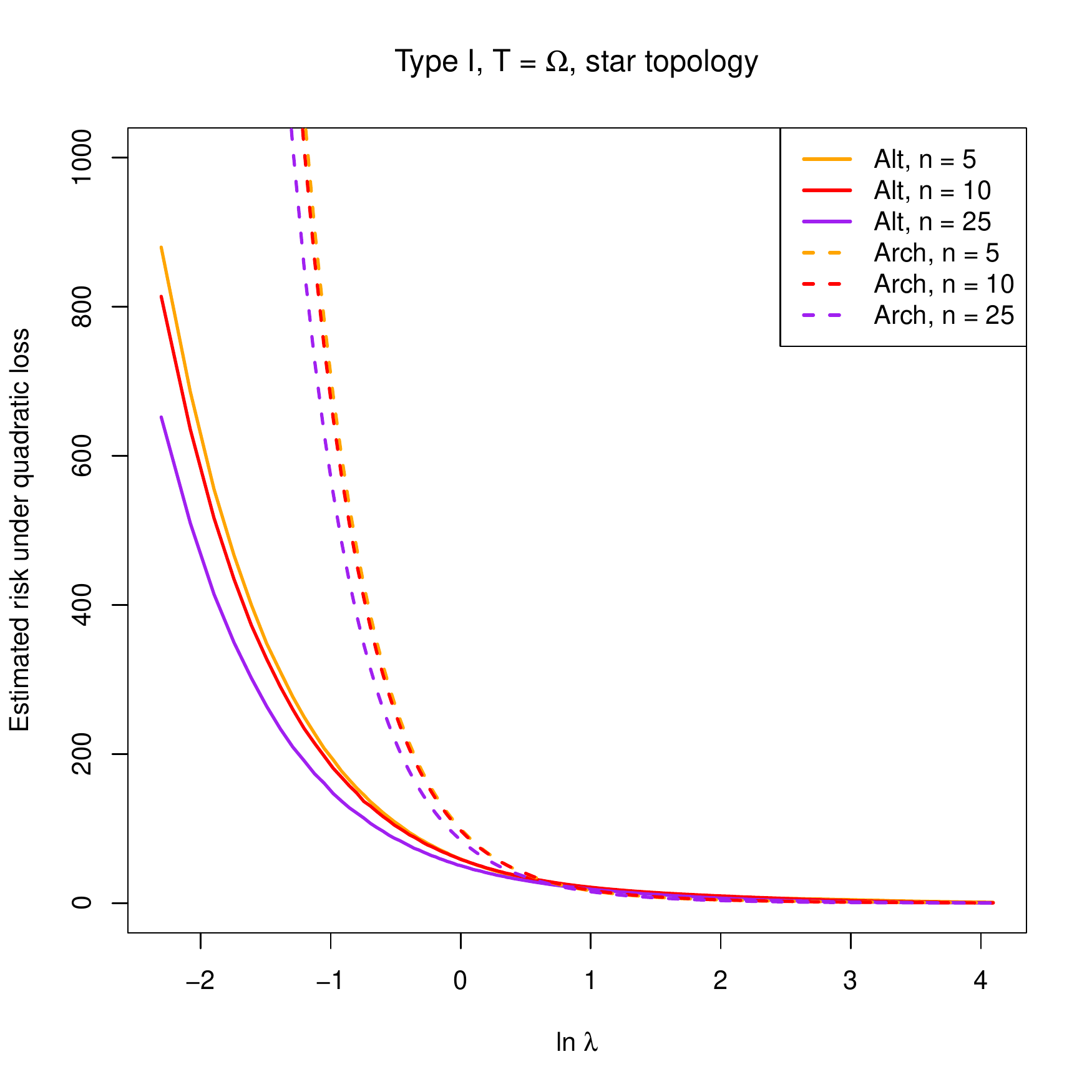}
    \caption{Estimated risk vs. penalty parameter. All panels display, for the star topology,
the estimated risks under quadratic loss for Type I ridge
estimators. The left panel compares the alternative and archetypal
Type I ridge estimators when the target is taken to be
\emph{$\mbox{\emph{diag}}[\mathbf{T}] = 1/\mbox{\emph{diag}}[\mathbf{S}]$}. The
right hand panel compares the alternative and archetypal Type I
ridge estimators when $\mathbf{T} = \mathbf{\Omega}$. The dashed
lines represent the archetypal estimator while the solid lines
represent the alternative estimator. The orange, red and purple line
colorings represent the various sample sizes ($n=5, 10, 25$,
respectively). Note that the fluctuations in the estimated risks in
the left-hand panel are due to the data dependency of the target.
Also note that, for purposes of comparability, the scales of the
$\lambda$ parameter under the various estimators were chosen in
accordance with the eigenvalue comparison in Section
\ref{ShrinkMyEigenValue}.}\label{RiskFig}
\end{figure}

The risk is evaluated on data sets drawn from a multivariate normal
distribution with four different (population) precision matrices:
\begin{enumerate}
\item $\mathbf{\Omega}^{\mbox{{\tiny random}}}$ with no conditional dependencies, generated as $\mathbf{\Omega}^{\mbox{{\tiny random}}}  =  \frac{1}{n}\mathbf{Y}^{\mathrm{T}} \mathbf{Y}$ from  the $(n \times p)$-dimensional matrix $\mathbf{Y}$ with $n=10,000$ and each $Y_{ij}$ drawn from $\mathcal{N}(0,1)$;\vspace{-.2cm}
\item $\mathbf{\Omega}^{\mbox{{\tiny chain}}}$ representing a conditional independence graph with a chain topology. Its element are $(\mathbf{\Omega}^{\mbox{{\tiny chain}}})_{j,j}  =  1$, $(\mathbf{\Omega}^{\mbox{{\tiny chain}}})_{j,j+1}  =  0.25 = (\mathbf{\Omega}^{\mbox{{\tiny chain}}})_{j+1,j}$ for $j=1, \ldots, p-1$, and zero otherwise;\vspace{-.2cm}
\item $\mathbf{\Omega}^{\mbox{{\tiny star}}}$ representing a conditional independence graph with a star topology. Its element are $(\mathbf{\Omega}^{\mbox{{\tiny star}}})_{j,j}  =  1$, $(\mathbf{\Omega}^{\mbox{{\tiny star}}})_{1,j+1}  =  1/(j+1) = (\mathbf{\Omega}^{\mbox{{\tiny star}}})_{j+1,1}$ for $j=1, \ldots, p-1$, and zero otherwise;\vspace{-.2cm}
\item $\mathbf{\Omega}^{\mbox{{\tiny clique}}}$ representing a conditional independence graph with a clique structure. The structure consists of five equally sized blocks along the diagonal, each with unit diagonal elements and off-diagonal elements equal to 0.25.
\end{enumerate}
Throughout the simulation the dimension of $p$ is fixed at $p=100$
while the sample size varies: $n=5, 10$ and $25$. This represents
varying degrees of high-dimensionality. For each combination of
precision matrix and sample size one hundred data sets are drawn.
For each draw the sample covariance matrix is calculated. The
penalized estimates of the precision matrix are obtained for a large
grid of the penalty parameter using the Type II null-matrix target
($\mathbf{T} = \mathbf{0}$), a diagonal target
$(\mbox{diag}[\mathbf{T}] = 1/\mbox{diag}[\mathbf{S}])$, and a
target equal to the true precision matrix $(\mathbf{T} =
\mathbf{\Omega})$. Note that in the comparison for the latter two
Type I situations the archetypal target $\mathbf{\Gamma}$ is taken
to be $\mathbf{T}^{-1}$, so that the archetypal and alternative
estimators have the same target in the precision sense. For each
penalized precision estimate the quadratic and Frobenius loss are
evaluated and subsequently the risk (under given loss function) is
approximated by the median loss over the hundred draws. Figure
\ref{RiskFig} shows, for the star topology, the estimated risks
under quadratic loss for Type I ridge estimators
($\mbox{diag}[\mathbf{T}] = 1/\mbox{diag}[\mathbf{S}]$ and
$\mathbf{T} = \mathbf{\Omega}$) plotted against the penalty
parameter (see Section 2 of the Supplementary Material for
visualizations of all risk comparisons).

The simulation results (as summarized in Figure \ref{RiskFig} and
Section 2 of the Supplement) show that the alternative Type I ridge
estimator outperforms its archetypal counterpart with respect to
both loss types (when shrinking towards either of the non-zero
targets). This behavior holds irrespective of the generated
population precision matrix, the $p/n$ ratio, and the choice of
target. The superior performance of the alternative Type I estimator
is strongest for small to medium-sized values of the penalty
parameter (this will correspond, in practice, to the most relevant
part of the domain). For large values of the penalty parameter the
loss difference vanishes. This due to the fact that both alternative
and archetype shrink to the same target. For both estimators the
spot-on target ($\mathbf{T} = \mathbf{\Omega}$) yields a lower loss
for large values of $\lambda$ than the diagonal target. The gain of
employing a spot-on target increases, as can be expected, with the
$p/n$ ratio. With regard to Type II estimation the estimated risks
of the alternative and archetypal estimators are similar, although
the alternative estimator performs marginally better. In all, the
alternative ridge precision estimators outperform their archetypal
counterparts in this simulation study.
%%%%%%%%%%
%%%%%%%%%%

%%%%%%%%%%
%%%%%%%%%%
\section{Comparing Alternative Ridge and Graphical Lasso Estimation}\label{Illustrate}
A contemporary use for precision matrices is found in network
reconstruction through graphical modeling. Graphical modeling refers
to a class of probabilistic models that uses graphs to express
conditional (in)dependence relations between random variables. In
this section we investigate how well the proposed ridge estimators
of the precision matrix uncover conditional (in)dependencies from
high-dimensional data. The performance of the alternative ridge
estimators is contrasted with the graphical lasso \citep{Frie2008};
the lasso estimator of the precision matrix. Two versions of each
estimator are considered. On the ridge side the Type II alternative
ridge precision estimator with $\mathbf{T} = \mathbf{0}$ and the
Type I alternative ridge estimator with $\mbox{diag}[\mathbf{T}] =
1/\mbox{diag}[\mathbf{S}]$ are considered. The concordant graphical
lasso precision estimators employ penalization and no penalization
of the diagonal elements, respectively \citep[see the \texttt{glasso}
package:][]{glassoMAN}. In order to avoid any bias towards either
method of estimation, the comparison makes use of real data while
adhering to the \emph{ceteris paribus} principle with regard to
penalty parameter selection (see also below). In the remainder of this
section we will first review graphical modeling (Section \ref{GraphModel})
and the data (Section \ref{DataGoalCompare}), before focusing the comparison
on loss (Section \ref{LC}), sensitivity and specificity (Section
\ref{Sensitive}), and network stability (Section \ref{Cigs}),
respectively.

\subsection{Graphical Modeling}\label{GraphModel}
We consider graphs $\mathcal{G} = (\mathcal{V}, \mathcal{E})$
consisting of a finite set $\mathcal{V}$ of vertices and set of
edges $\mathcal{E}$. The vertices of the graph correspond to a
collection of random variables with probability distribution
$\mathcal{P}$, i.e., $\{Y_{1},\ldots,Y_{p}\} \sim \mathcal{P}$.
Edges in $\mathcal{E}$ consist of pairs of distinct vertices such
that $Y_{j} - Y_{j'} \in \mathcal{E}$. The basic assumption is:
$\{Y_{1},\ldots,Y_{p}\} \sim \mathcal{N}_{p}(\boldsymbol{0},
\mathbf{\Sigma})$, with $\mathbf{\Sigma} \succ 0$. We thus focus on
Gaussian graphical modeling by considering pairs $(\mathcal{G},
\mathcal{P} \sim \mathcal{N})$. (See Figure \ref{GGMs} for a visual
example of a graphical model).

In this Gaussian case, conditional independence between a pair of
variables corresponds to zero entries in the precision matrix.
Indeed, let $\hat{\mathbf{\Omega}}$ denote a generic estimate of the
precision matrix and consider its transformation to a partial
correlation matrix $\hat{\mathbf{P}}$. Then the following relations
can be shown to hold for all pairs $\{Y_{j}, Y_{j'}\} \in
\mathcal{V}$ with $j \neq j'$ \citep[see, e.g.,][]{whittaker}:
\begin{equation}\nonumber
(\hat{\mathbf{P}})_{jj'} = 0
\Longleftrightarrow (\hat{\mathbf{\Omega}})_{jj'} = 0
\Longleftrightarrow Y_{j} \ci Y_{j'}|\mathcal{V}\setminus\{Y_{j},Y_{j'}\}
\Longleftrightarrow Y_{j} \centernot{-} Y_{j'},
\end{equation}
where $\mathcal{V}\setminus\{\cdot\}$ denotes set-minus notation and
where $\centernot{-}$ indicates the absence of an edge. Hence, model
selection efforts in Gaussian graphical models focus on determining
the support of the precision matrix.

The graphical lasso \citep{Frie2008} performs, next to shrinkage,
automatic selection of conditional dependencies. As the alternative
ridge estimators will not generally produce sparse estimates, they
will need to rely on an additional procedure for support
determination. Here, we resort to a multiple testing procedure.
Specifically, we use the local false discovery rate (lFDR) procedure
\citep{EfronLocFDR} proposed by \cite{SS05}. Let
$\hat{\mathbf{P}}^{\mathrm{I}a}(\lambda_{a})$ denote the regularized
precision estimate
$\hat{\mathbf{\Omega}}^{\mathrm{I}a}(\lambda_{a})$ scaled to partial
correlation form. For support determination, we assume that the
nonredundant off-diagonal partial correlation coefficients (indexed
by, say, $j < j'$) follow a mixture distribution:
\begin{equation}\nonumber
f\left\{ [\hat{\mathbf{P}}^{\mathrm{I}a}(\lambda_{a})]_{jj'} \right\} =
\eta_{0} f_{0} \left\{ [\hat{\mathbf{P}}^{\mathrm{I}a}(\lambda_{a})]_{jj'} ; \kappa \right\} +
(1 - \eta_{0}) f_{\mathcal{E}} \left\{ [\hat{\mathbf{P}}^{\mathrm{I}a}(\lambda_{a})]_{jj'} \right\},
\end{equation}
with mixture weight $\eta_{0} \in [0,1]$, and where $f_{0}\{\cdot\}$
denotes the distribution of a null-edge while
$f_{\mathcal{E}}\{\cdot\}$ denotes the distribution of a present
edge. The former density can be found to be a scaled beta-density
\citep{Hotel, SS052, SS05}:
\begin{equation}\nonumber
f_{0} \left\{ [\hat{\mathbf{P}}^{\mathrm{I}a}(\lambda_{a})]_{jj'} ; \kappa \right\} =
\left| [\hat{\mathbf{P}}^{\mathrm{I}a}(\lambda_{a})]_{jj'} \right|
\mathcal{B}\left\{ [\hat{\mathbf{P}}^{\mathrm{I}a}(\lambda_{a})]_{jj'}^{2}; \frac{1}{2}, \frac{\kappa - 1}{2} \right\},
\end{equation}
with $\kappa$ degrees of freedom (note that in the last expression
$|\cdot|$ is used to denote the absolute value). In the $p > n$
situation $\kappa$ has to be estimated, next to $\eta_{0}$ and
$f_{\mathcal{E}}\{\cdot\}$. See \cite{EfronLSbook} and \cite{SS05}
for details on obtaining estimates of these unknowns. Having these
estimates at hand, the lFDR is given as \citep{SS05}:
\begin{equation}\nonumber
P\left(Y_{j} \centernot{-} Y_{j'} | [\hat{\mathbf{P}}^{\mathrm{I}a}(\lambda_{a})]_{jj'} \right) =
\frac{\hat{\eta}_{0} f_{0} \left\{ [\hat{\mathbf{P}}^{\mathrm{I}a}(\lambda_{a})]_{jj'} ; \hat{\kappa} \right\}}
{\hat{\eta}_{0} f_{0} \left\{ [\hat{\mathbf{P}}^{\mathrm{I}a}(\lambda_{a})]_{jj'} ; \hat{\kappa} \right\} +
(1 - \hat{\eta}_{0}) \hat{f}_{\mathcal{E}} \left\{ [\hat{\mathbf{P}}^{\mathrm{I}a}(\lambda_{a})]_{jj'} \right\}},
\end{equation}
conveying the empirical posterior probability that the edge between
$Y_{j}$ and $Y_{j'}$ is null given
$[\hat{\mathbf{P}}^{\mathrm{I}a}(\lambda_{a})]_{jj'}$. Another
useful quantity is $1 - \mbox{lFDR}$, indicating the analogous
probability that an edge is present. Again, similar probabilistic
statements can be made with the alternative Type II estimator when
replacing in the above $\hat{\mathbf{P}}^{\mathrm{I}a}(\lambda_{a})$
by $\hat{\mathbf{P}}^{\mathrm{II}a}(\lambda_{a})$. In Sections \ref{Sensitive}
and \ref{Cigs} an edge will be selected when $1 - \mbox{lFDR} \geq .99$.

While the two-step procedure of regularization followed by
subsequent support determination does not have the appeal of
simultaneous estimation and model selection, it does have the
advantage that it enables probabilistic statements about the
inclusion (or exclusion) of edges. An additional advantage is that
the procedure may lead to a better representation of individual
partial correlation or precision elements after sparsification: The
closest, in a least-squares sense, p.d. sparsified representation of
$\hat{\mathbf{\Omega}}^{\mathrm{I}a}(\lambda_{a})$ (or
$\hat{\mathbf{\Omega}}^{\mathrm{II}a}(\lambda_{a})$), is indeed
$\hat{\mathbf{\Omega}}^{\mathrm{I}a}(\lambda_{a})$ (or
$\hat{\mathbf{\Omega}}^{\mathrm{II}a}(\lambda_{a})$) with the
zero-structure imposed as follows from the lFDR test \citep[cf.][]{LSadjust}.

\subsection{Data}\label{DataGoalCompare}
The performance of the ridge and lasso precision estimators is
evaluated on gene expression data of three pathways from five
oncogenomics studies. The Bioconductor repository \citep{BioCon}
offers five curated breast cancer data sets \citep{BreastData}
generated on the same microarray platform (Affymetrix hgu 133
platform). These datasets will be indicated as follows: Mainz,
Transbig, UNT, UPP, VDX. The data of these studies have been
preprocessed in a uniform manner \citep[see][]{QuackB12}.
Cancer of the breast is a hormone-related cancer, with a central
role for estrogen. Breast cancerous tissue may have many estrogen
receptors (ER+ breast cancer) or few estrogen receptors (ER$-$ breast
cancer). The genomic pathways of ER+ and ER$-$ breast cancers differ.
Thus, to remove further heterogeneity among the data sets, they are
limited to ER+ samples. The chosen pathways, p53, apoptosis, and
mTOR, are defined by KEGG \citep{KEGG}. The p53 gene is a tumor
suppressor gene. Cellular stress signals such as DNA damage can
activate the p53-pathway, resulting in a multilayered tumor
suppressive mechanism \citep{MBC}. The genetic p53-pathway is defined
to consist of those genes mediating the path from cellular stress
signal to p53-induced tumor suppressive response. Alterations of the
p53 pathway are found in most human cancers \citep{Vogelp53}.
Apoptosis refers to the process of regulated cell death. The ability
of cancerous cells to resist apoptosis is considered to be one of
the hallmarks of human cancer \citep{Hallmarks}. The mTOR protein is
a kinase (a phosphate transferring enzyme) that is frequently
overexpressed in human cancers. This may lead to oncogenic
signaling, making the cancerous cell self-sufficient in survival and
multiplication \citep{MBC}, another hallmark of human cancer
\citep{Hallmarks}. The underlying conditional dependency structure of
the respective pathways is not fully known but is (generally)
believed to be (relatively) sparse.

For each data set the probe sets that interrogate genes mapping to
the p53, apoptosis, and mTOR pathways are selected. Whenever
multiple probe sets map to the same gene, their expression levels
have been averaged sample-wise over the instances. The resulting
dimensions of the $n \times p$ pathway data sets are: $n=162$
(Mainz), $n=134$ (Transbig), $n=86$ (UNT), $n=213$ (UPP), $n=209$
(VDX), and $p=67$ (p53), $p=83$ (apoptosis), $p=47$ (mTOR). See
Section 3 of the Supplement for \texttt{R} code on extracting the
mentioned data.

The pathway data are not high-dimensional in the sense $p > n$.
High-dimensionality is achieved by subsampling with sample sizes $n
= 5, 10$ and $25$. One hundred subsamples are drawn of each
mentioned sample size for each combination of pathway and breast
cancer data set. Optimal values of the penalty parameter for both
versions of the alternative ridge and lasso estimators are obtained
for each subsample by way of LOOCV. The ridge and lasso precision
estimates for a subsample then correspond to these optimal penalty
parameter values. Finally, the estimates are standardized to have
unit diagonal (the standardized precision matrix is equal to the
partial correlation matrix up to the sign of off-diagonal entries).

\subsection{Loss Comparison}\label{LC}
The standardized precision estimates are evaluated in terms of
quadratic and Frobenius loss (as defined in Section \ref{RC}). This
requires the standardized population precision matrix, which is
unknown. As a proxy we take the sample version obtained from the
data with all samples, e.g., the standardized population precision
matrix for the p53-pathway in the UPP data is defined as the $(67
\times 67)$-dimensional standardized sample precision matrix over
all $n=213$ samples. The results of the loss evaluation are
displayed in Figure \ref{gRidgeLasso} and Section 4 of the
Supplementary Material.

Figure \ref{gRidgeLasso} and Section 4.1 of the Supplement show that
the quadratic loss of the lasso estimate of the standardized
precision matrix exceeds that of its ridge counterpart. In general,
this is a consistent observation over the sample sizes ($n=5, 10$
and $25$), the pathways, and the data sets. This behavior also holds
for the Frobenius loss and holds irrespective of the choice of
target. In several cases the loss difference between the estimators
decreases as $n$ increases. This should not surprise, as the loss
difference is expected to vanish for large $n$ under fixed $p$ (also
note that, naturally, loss decreases with increasing $n$). Thus, the
alternative ridge estimators of the standardized precision matrix
yield a lower loss than the corresponding lasso estimators, in
particular for the larger $p/n$ ratios.

\begin{figure}[h]
\begin{center}
\begin{tabular}{cc}
\includegraphics[scale=0.4]{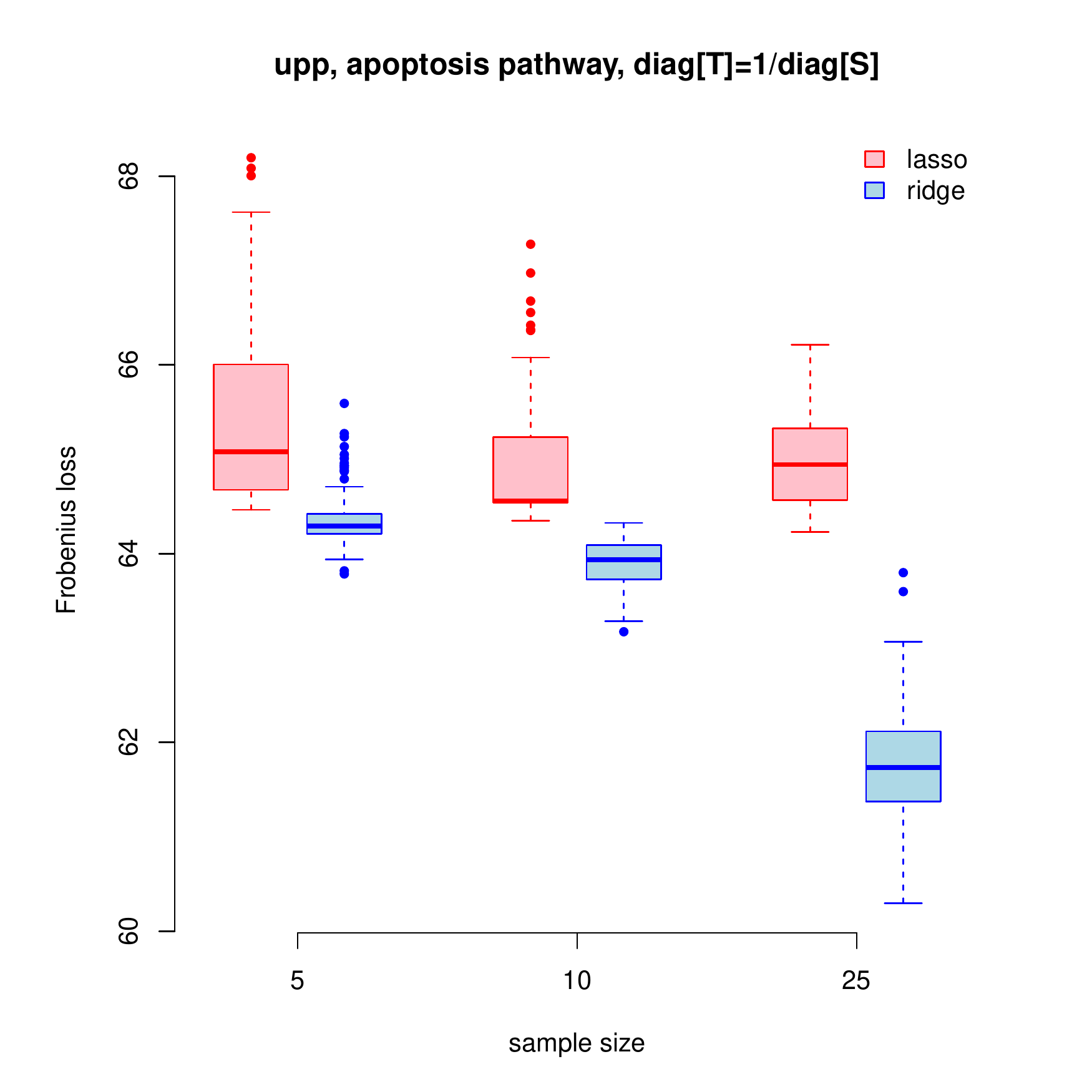}
& \hspace{-1cm}
\includegraphics[scale=0.4]{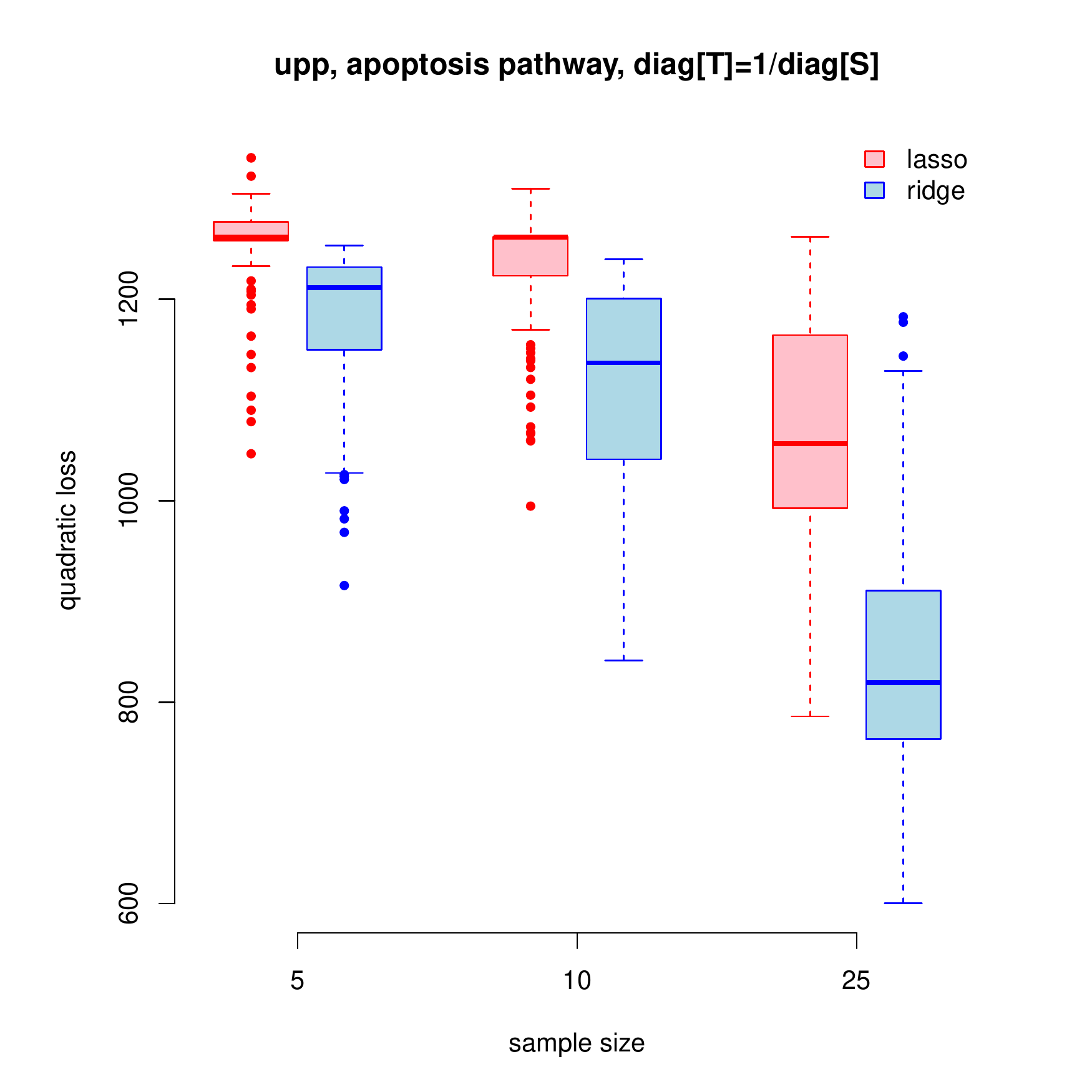}
\end{tabular}
\caption{Loss comparison between the Type I alternative ridge
estimator with $\mbox{diag}[\mathbf{T}] =
1/\mbox{diag}[\mathbf{S}]$ and the corresponding graphical lasso
estimator on the UPP apoptosis-pathway data. The left-hand panel depicts Frobenius loss while the
right-hand panel depicts quadratic loss.}\label{gRidgeLasso}
\end{center}
\end{figure}

\subsection{Sensitivity and Specificity}\label{Sensitive}
The evaluation of sensitivity and specificity of edge selection
requires knowledge of the true conditional dependencies. Such
knowledge is absent as the (causal) biological mechanisms underlying
the pathway are mostly unknown (or at least uncertain). Hence, we
resort to defining a `consensus truth', comprised of those
conditional dependencies that appear in the top $100 \alpha \%$ of
at least 4 out of the 5 breast cancer data sets by both methods
(graphical lasso and alternative ridge paired with lFDR edge
selection). The top $100 \alpha \%$ constitutes of the $\lceil
\frac{1}{2} p (p-1) \alpha \rceil$ edges with the largest selection
frequency over the hundred respective subsamples (see Section
\ref{DataGoalCompare}), with $\alpha = \{0.005, 0.01, 0.015, \ldots,
0.20\}$. This yields a nested sequence of `consensus truths'. The
range of $\alpha$ corresponds to what is believed to be biologically
plausible. Thus, with observed selected edges and the `consensus
truths' at hand, sensitivity and specificity are estimated per
subsample over the range of $\alpha$. The median sensitivity
(specificity) over the hundred subsamples over all data sets is
taken as the estimate of the sensitivity (specificity) for a
particular combination of $\mathbf{T}$, $n$, $\alpha$, and pathway.
Figure \ref{SeSpRidgeLasso} and Section 4.2 of the Supplementary
Material visualize estimated sensitivity and specificity against
$\alpha$.

\begin{figure}[b!]
\begin{center}
\includegraphics[scale=0.4]{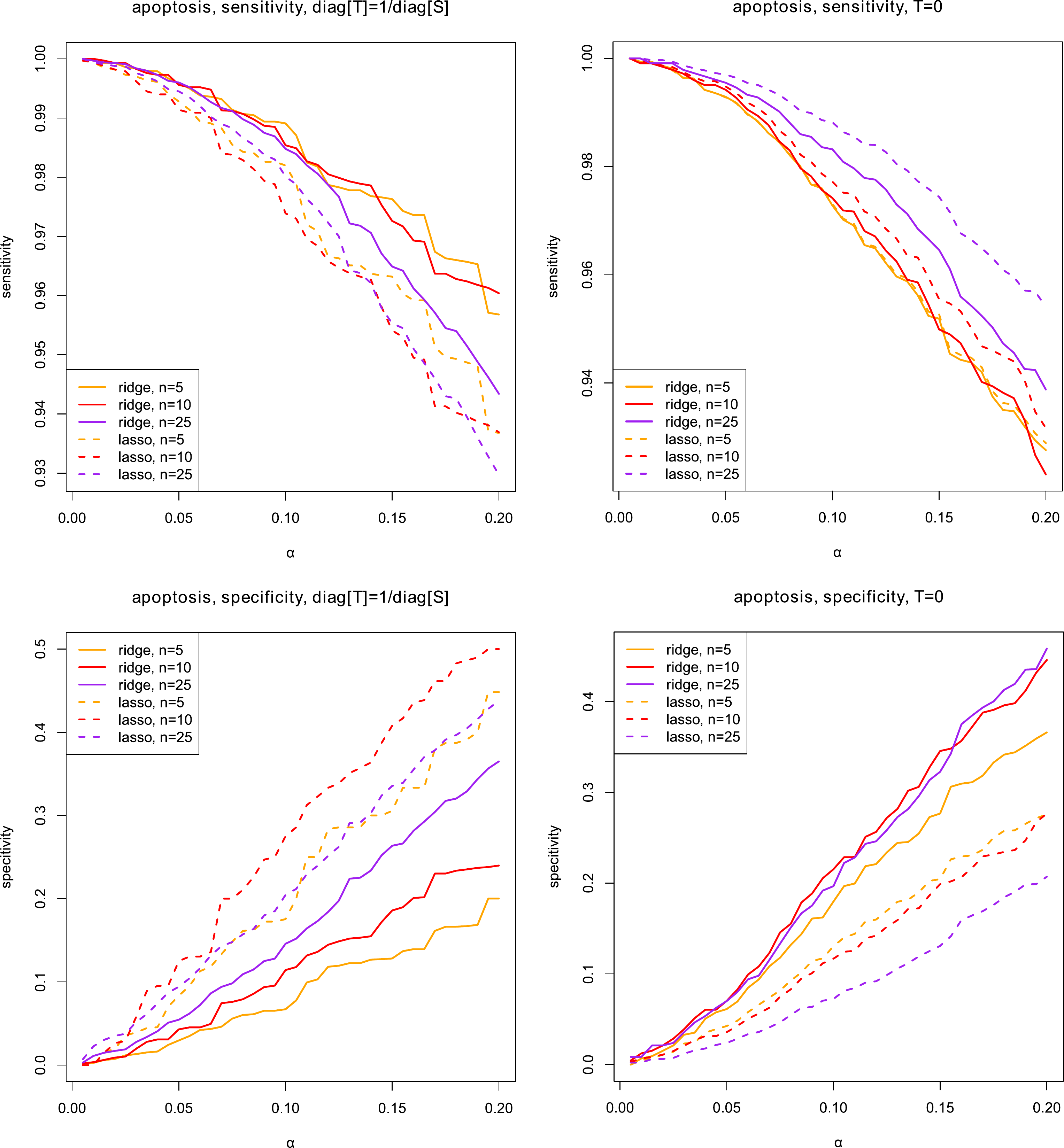}
\caption{Sensitivity and specificity comparison between the
alternative ridge and graphical lasso estimators on the
apoptosis-pathway data. The upper panels depict sensitivity results
while the lower panels depict specificity results. The left-hand
panels depict results for $\mbox{diag}[\mathbf{T}] =
1/\mbox{diag}[\mathbf{S}]$ while the right-hand panels depict
results for $\mathbf{T} = \boldsymbol{0}$.}\label{SeSpRidgeLasso}
\end{center}
\end{figure}

First note that the sensitivity is high for both the ridge estimator
and the graphical lasso (in both Type I and Type II situations), due
to the stringent definition of `consensus truth'. Furthermore, the
Type I ridge estimator outperforms the corresponding lasso in terms
of sensitivity. For the Type II setting it is seen that the
graphical lasso fares slightly better with regard to sensitivity.
These behaviors are reversed when evaluating specificity: The lasso
fares better in the Type I setting, while the ridge outperforms the
lasso in the Type II situation. These observations hold for all
pathways. These findings can (at least in part) be traced to the
utilization of lFDR edge selection on the ridge estimators (cf.
Section \ref{GraphModel}). Using $\mathbf{T} = \boldsymbol{0}$ will
(tend to), by enforcing more uniformity among the partial
correlation values, emphasize the null-edge distribution, leading to
improved specificity and (somewhat) diminished sensitivity. A p.d.
target $\mathbf{T}$, on the other hand, will tend to preserve data
signal, and will subsequently lead to improved sensitivity and
(somewhat) diminished specificity. These behaviors might suggest the
following (also taking into account the loss behavior and the
stability of performance over respective sample sizes): Give
preference to the Type I alternative ridge estimator when
emphasizing the true positive rate, and give preference to the Type
II alternative ridge estimator when emphasizing the true negative
rate.

\subsection{Stability}\label{Cigs}
The performance of the (Type I and II) ridge and lasso precision
estimators can also be evaluated in terms of network stability.
Define an edge stable when it is selected in the union of the top
$100 \alpha \%$ over the respective subsample sizes $n = 5, 10$, and
$25$.  When plotting the number of stable edges against $\alpha$
(see Figure \ref{EdgeStability} and Section 4.3 of the Supplement),
it is clear that the number of stable edges shows a faster increase,
with increasing $\alpha$, for the ridge estimators than for the
graphical lasso. This effect is especially pronounced for the Type I
ridge setting. The ridge estimators also sort more stable behavior
over the respective data sets.

\begin{figure}[t!]
    \centering
    \includegraphics[scale=.42]{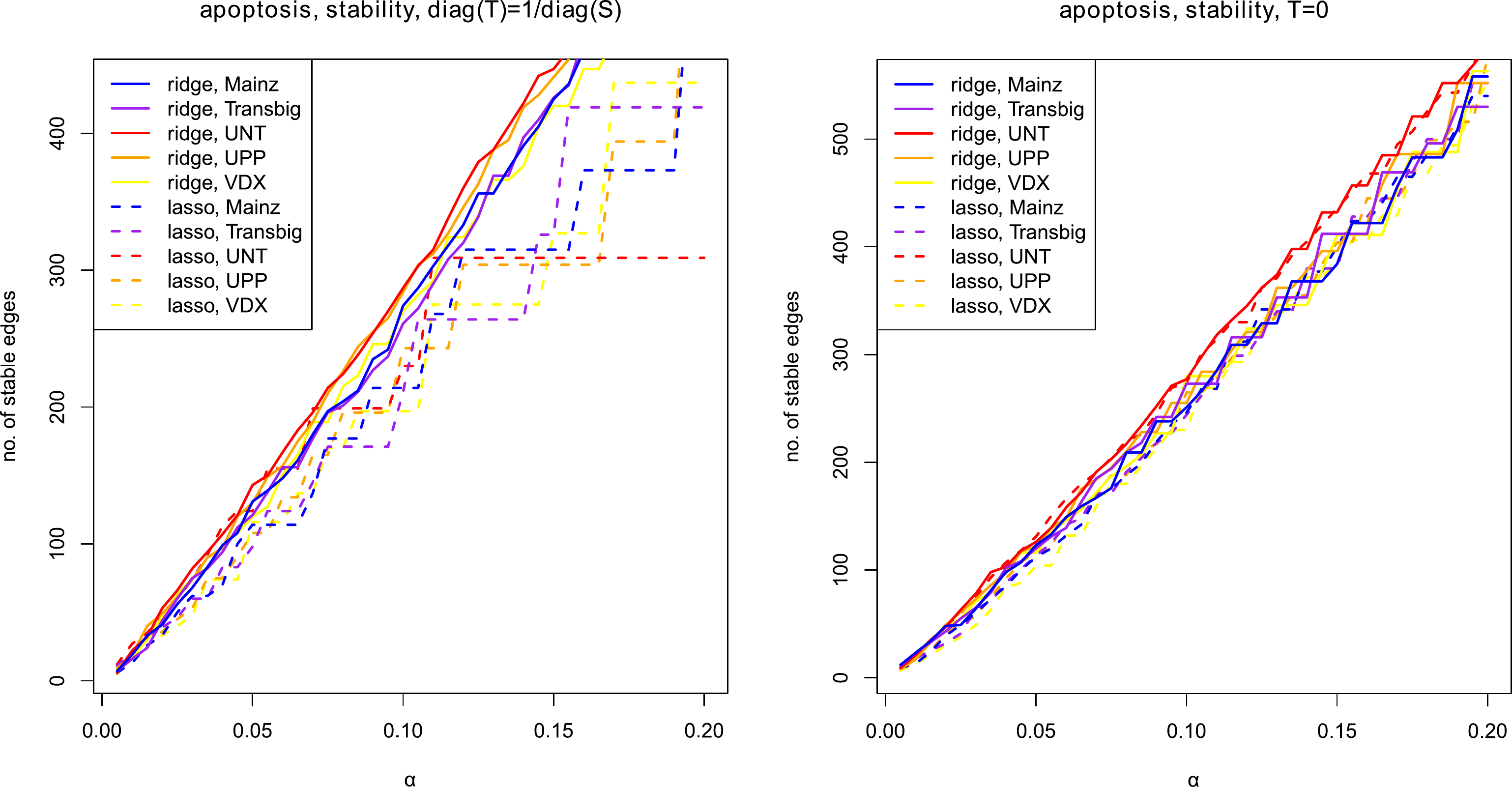}
    \caption{The number of stable edges plotted against $\alpha$ for the apoptosis-pathway data.
The left-hand panel depicts results for
$\mbox{diag}[\mathbf{T}] = 1/\mbox{diag}[\mathbf{S}]$ while
the right-hand panel depicts results for $\mathbf{T} =
\boldsymbol{0}$.}\label{EdgeStability}
\end{figure}

\begin{figure}[t!]
    \centering
    \includegraphics[scale=.36]{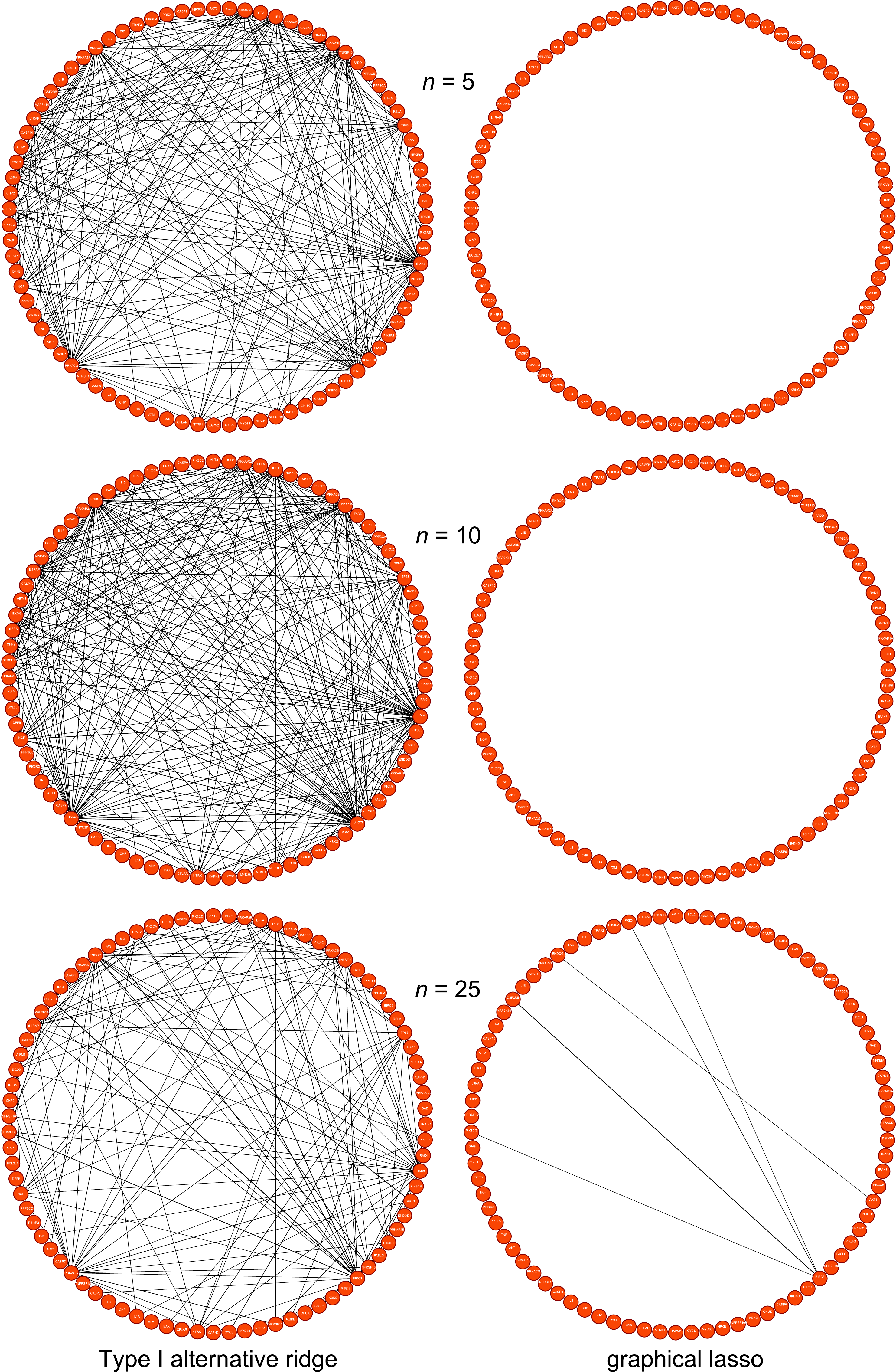}
    \caption{Conditional independence graphs for the Type I alternative
ridge estimator using lFDR edge selection (left-hand figures)
and the corresponding graphical lasso (right-hand figures) on the
UPP apoptosis-pathway data. For an edge to be represented in the
conditional independence graphs above, it must have been selected at
least 50 times over the 100 replications (given sample size $n=5,
10$ and $25$, respectively).}\label{GGMs}
\end{figure}

Analogous behavior can be shown with regard to the effect of sample size.
Figure \ref{GGMs} contains conditional independence graphs for the
Type I alternative ridge estimator with $\mbox{diag}[\mathbf{T}] =
1/\mbox{diag}[\mathbf{S}]$ and the corresponding graphical lasso on
the UPP apoptosis-pathway data. A represented edge means that it
was selected at least 50 times over the 100 subsamples. It may be
observed that the pairing of the alternative ridge estimator with
lFDR support determination selects more stable
(in terms of network-structure change) networks over
the respective sample sizes. While usage of lFDR edge selection
on the ridge regularized precision matrix tends to gain in
conservativeness with growing $n$,
the network-structure changes over the respective sample sizes are
much less dramatic vis-\`{a}-vis the graphical lasso. This picture
of stability holds for the remaining pathways
under $\mbox{diag}[\mathbf{T}] = 1/\mbox{diag}[\mathbf{S}]$. For the Type II comparison the
alternative ridge estimator tends to be more conservative than the
graphical lasso for the higher sample sizes (cf. explanation Section \ref{Sensitive}).
This can again be taken as an indication that
when the network data at disposal do contain a sizeable signal, it
is preferable to choose a non-null target $\mathbf{T}$ for better
signal preservation.

\subsection{The Graphical Lasso as Reference}\label{LassoLead}
One may argue that the comparability may be obscured when the number
of selected edges differs considerably between methods. We thus, in
addition to the exercises above, take interest in comparing the
alternative ridge estimators with the graphical lasso when the
latter dictates the number of edges the former may select. Say the
graphical lasso selects, within a certain subsample, $\tau$ edges;
then for the corresponding ridge precision estimator the $\tau$
edges are selected with the largest absolute partial correlation. It
is obvious that thresholding the ridge precision estimator on the
basis of the graphical lasso will favor the latter. In this setup
the ridge estimators thus prove their strength through
non-inferiority. The results in Section 5 of the Supplementary
Material show that, indeed, the alternative ridge estimators display
non-inferiority with respect to sensitivity, specificity, and
stability in this situation.

Summarizing on the basis of the results in Sections
\ref{LC}--\ref{LassoLead}: The alternative Type I (Type II) ridge
estimator paired with post-hoc edge selection is a contender in a
graphical modeling setting, especially when the $p/n$ ratio tends to
get more extreme and/or when emphasis is placed on the true positive
(negative) rate.
%%%%%%%%%%
%%%%%%%%%%

%%%%%%%%%%
%%%%%%%%%%
\section{Discussion}\label{Discuss}
We studied ridge estimation of the precision matrix. Estimators
currently in use can be roughly divided into two archetypes whose
penalties do not coincide with the common ridge penalty. Starting
from the common ridge penalty we derived an analytic expression of
the ridge estimator of the inverse covariance matrix, on the basis
of which alternatives were formulated for the two archetypes. The
alternative estimators were shown to outperform the archetypes in
terms of risk. An illustration using pathway data also showed that
the alternative ridge estimators perform better than the
corresponding graphical lasso estimators in terms of loss. They also
tend to select more stable networks, especially in situations where
the variable to sample ratio is more extreme. The provided
expressions can also be of use in the study of theoretical
properties of penalized inverse covariance estimators.

The proposed estimators can facilitate methods and approaches of
data analysis leaning on the estimation of precision (or covariance)
matrices in high-dimensional situations. For example, the estimators
may be used in supporting covariance regularized regression
\citep{Wit09}, discriminant analysis, or canonical correlation
analysis. In addition, in the context of graphical modeling, the
proposed estimators can be paired with post-hoc methods for
determining the support of the precision matrix, such as lFDR
multiple testing \citep{SS05}. Furthermore, regularized (inverse)
covariance matrices stemming from the proposed estimators can be
used as input in covariance structure modeling efforts \citep{Jor81}
(including factor analysis and structural equation modeling as
special cases), when $p$ is large relative to $n$.

We see various inroads for further research. One would be to study
the proposed estimators from a Bayesian perspective. In addition,
the Type I estimator may lend itself for a natural framework of
Bayesian updating regarding graphical modeling, where the target is
determined by previous rounds of fitting the estimator followed by
subsequent support determination. Another option would be to extend
the proposed estimators with a condition number constraint
\citep{Won2013}, so that it can be formalized which values for the
penalty parameter can be considered `too small'. Also, the results
from the numerical studies may be further supported with results on
increasing-dimension asymptotics of the proposed estimators. From a more applied
perspective it may be deemed interesting to compare multiple post-hoc methods
for determining the support of the precision matrix. These
issues are the focal points of current research.

\begin{sloppypar}
The ridge estimators employed in this paper are implemented in the
\texttt{R}-package \texttt{rags2ridges} \citep{rags} along with supporting functions to
employ these estimators in a graphical modeling setting. The package
is freely available from the Comprehensive \texttt{R} Archive Network
(\url{http://cran.r-project.org/}) \citep{Rman}.
\end{sloppypar}

%---- Acknowledgements
%-----------------------------------------------------------------------------------
\section*{Acknowledgements}
The research leading to these results has received funding
from the European Community's Seventh Framework Programme (FP7,
2007-2013), Research Infrastructures action, under the grant
agreement No. FP7-269553 (EpiRadBio project). The authors would also
like to thank Mark van de Wiel, Poul Svante Eriksen, and Gr\'{e}gory
Nuel, whose constructive comments have
led to an improvement in presentation.

%---- Appendices
%-----------------------------------------------------------------------------------
\appendix
\section{Proofs}\label{Proofs}
This appendix contains proofs for Lemma \ref{GenRidgeAltLemma}, Propositions \ref{RidgeAltIProp}, \ref{prop.consistency}, \ref{AltIEIGprop} and \ref{AltIIEIGprop}, as well as Corollaries \ref{RidgeAltIIColl} and \ref{AltIILIKEcor}.
Consider first the following theorem:

\begin{theorem}\emph{\citep[p.~115]{Serre02}}\label{SQRt}
Let $\mathbf{H}$ be a p.d. Hermitian matrix. There exists a unique p.d. Hermitian matrix $h$ such that
$h^2 = \mathbf{H}$. If $\mathbf{H}$ is real-valued, then so is $h$. The matrix $h$ is called the \emph{square root}
of $\mathbf{H}$, and is denoted by $h = \mathbf{H}^{1/2}$.
\end{theorem}
This theorem is of use in the proof of Lemma \ref{GenRidgeAltLemma}:

\bigskip
\newproof{poL1}{Proof of Lemma \ref{GenRidgeAltLemma}}
\begin{poL1}\label{GenRidgeAltLemmaProof}
Define the penalized log-likelihood:
\begin{equation}\nonumber
\mathcal{L}^{p}(\mathbf{\Omega};\mathbf{S}, \mathbf{T}, \lambda_{a}) \propto \ln|\mathbf{\Omega}| - \mbox{tr}(\mathbf{S \Omega}) - \frac{\lambda_{a}}{2}\mbox{tr}[(\mathbf{\Omega}-\mathbf{T})^{\mathrm{T}}(\mathbf{\Omega}-\mathbf{T})].
\end{equation}
Now, take the derivative of $\mathcal{L}^{p}(\mathbf{\Omega};\mathbf{S}, \mathbf{T}, \lambda_{a})$ w.r.t. $\mathbf{\Omega}$:
\begin{align}\label{derivative}\nonumber
\frac{\partial \, \mathcal{L}^{p}(\mathbf{\Omega};\mathbf{S}, \mathbf{T}, \lambda_{a})}{\partial \, \mathbf{\Omega}} & =
2\left[\mathbf{\Omega}^{-1} - (\mathbf{S} - \lambda_{a}\mathbf{T}) - \lambda_{a}\mathbf{\Omega}\right] -
\left[\mathbf{\Omega}^{-1} - (\mathbf{S} - \lambda_{a}\mathbf{T}) - \lambda_{a}\mathbf{\Omega}\right] \circ \mathbf{I}_{p} \\
& = \left[\mathbf{\Omega}^{-1} - (\mathbf{S} - \lambda_{a}\mathbf{T}) - \lambda_{a}\mathbf{\Omega}\right] \circ
\left( 2\mathbf{J}_{p} -  \mathbf{I}_{p} \right),
\end{align}
where $\mathbf{J}_{p}$ denotes the all-ones matrix. It is immediate that (\ref{derivative}) is $\boldsymbol{0}$ only when
\begin{equation}\label{derivative0}
\mathbf{\Omega}^{-1} - (\mathbf{S} - \lambda_{a}\mathbf{T}) - \lambda_{a}\mathbf{\Omega} = \boldsymbol{0}.
\end{equation}
We will approach the problem in (\ref{derivative0}) from a square-completion angle.

Post-multiply (\ref{derivative0}) by $\mathbf{\Omega}^{-1}$. Subsequently adding $\frac{1}{4}(\mathbf{S} - \lambda_{a}\mathbf{T})^2$ to both sides of the equality sign gives that $\mathbf{\Omega}$ must satisfy:
\begin{eqnarray}\label{Ident1}
\lambda_{a}\mathbf{I}_{p} + \frac{1}{4}(\mathbf{S} - \lambda_{a}\mathbf{T})^2 := \mathbf{\Omega}^{-2} - (\mathbf{S} - \lambda_{a}\mathbf{T})\mathbf{\Omega}^{-1} + \frac{1}{4}(\mathbf{S} - \lambda_{a}\mathbf{T})^2.
\end{eqnarray}
Notice that under pre-multiplication by $\mathbf{\Omega}^{-1}$ the matrix $\mathbf{\Omega}$ is also implied to satisfy:
\begin{eqnarray}\label{Ident2}
\lambda_{a}\mathbf{I}_{p} + \frac{1}{4}(\mathbf{S} - \lambda_{a}\mathbf{T})^2 := \mathbf{\Omega}^{-2} - \mathbf{\Omega}^{-1}(\mathbf{S} - \lambda_{a}\mathbf{T}) + \frac{1}{4}(\mathbf{S} - \lambda_{a}\mathbf{T})^2.
\end{eqnarray}
Adding (\ref{Ident1}) and (\ref{Ident2}) and subsequently dividing by 2 thus yields:
\begin{eqnarray*}
\lambda_{a}\mathbf{I}_{p} + \frac{1}{4}(\mathbf{S} - \lambda_{a}\mathbf{T})^2 = \mathbf{\Omega}^{-2} - \frac{1}{2}\mathbf{\Omega}^{-1}(\mathbf{S} - \lambda_{a}\mathbf{T}) - \frac{1}{2}(\mathbf{S} - \lambda_{a}\mathbf{T})\mathbf{\Omega}^{-1} + \frac{1}{4}(\mathbf{S} - \lambda_{a}\mathbf{T})^2.
\end{eqnarray*}
Now, complete the square to obtain:
\begin{eqnarray}\label{CompSquare}
\lambda_{a}\mathbf{I}_{p} + \frac{1}{4}(\mathbf{S} - \lambda_{a}\mathbf{T})^2 = \left[\mathbf{\Omega}^{-1} - \frac{1}{2}(\mathbf{S} - \lambda_{a}\mathbf{T})\right]^2.
\end{eqnarray}
The left-hand side of (\ref{CompSquare}) is p.d., which implies that the right-hand side is p.d. By Theorem \ref{SQRt}, both sides then have a unique square root that is p.d. and symmetric. Taking this square root on both sides results in:
\begin{eqnarray}\nonumber
\left[\lambda_{a}\mathbf{I}_{p} + \frac{1}{4}(\mathbf{S} - \lambda_{a}\mathbf{T})^2\right]^{1/2} = \mathbf{\Omega}^{-1} - \frac{1}{2}(\mathbf{S} - \lambda_{a}\mathbf{T}).
\end{eqnarray}
Finally, solving for $\mathbf{\Omega}$ gives the desired expression (\ref{RidgeAltI}). \QEDE
\end{poL1}

%\bigskip
\newproof{poP1}{Proof of Proposition \ref{RidgeAltIProp}}
\begin{poP1}\label{RidgeAltIPropProof}
~

(i) Let $d(\cdot)_{jj}$ denote the $j$'th eigenvalue of the matrix term in brackets $(\cdot)$. Then
\begin{eqnarray*}
d\left\{[\hat{\mathbf{\Omega}}^{\mathrm{I}a}(\lambda_{a})]^{-1}\right\}_{jj} =
d\left[\frac{1}{2}(\mathbf{S} - \lambda_{a}\mathbf{T})\right]_{jj}
+
\sqrt{\left\{d\left[\frac{1}{2}(\mathbf{S} - \lambda_{a}\mathbf{T})\right]_{jj}\right\}^2
+ \lambda_{a}}
\, \, \, > \, \, \, 0,
\end{eqnarray*}
when $\lambda_{a} > 0$. Hence, $\hat{\mathbf{\Omega}}^{\mathrm{I}a}(\lambda_{a})$ is p.d. for any $\lambda_{a} \in (0,\infty)$.

(ii) The right-hand limit is immediate as:
\begin{eqnarray*}
\hat{\mathbf{\Omega}}^{\mathrm{I}a}(0) = \left\{\left[0\mathbf{I}_{p} + \frac{1}{4}(\mathbf{S} - 0\mathbf{T})^{2}\right]^{1/2} + \frac{1}{2}(\mathbf{S} - 0\mathbf{T})\right\}^{-1} = \mathbf{S}^{-1}.
\end{eqnarray*}

(iii) For the left-hand limit we note that, when $\lambda_{a}$ approaches $\infty$,
\begin{eqnarray*}
[\hat{\mathbf{\Omega}}^{\mathrm{I}a}(\lambda_{a})]^{-1} =
\left[\lambda_{a}\mathbf{I}_{p} + \frac{1}{4}(\mathbf{S} -
\lambda_{a}\mathbf{T})^{2}\right]^{1/2} +
\frac{1}{2}(\mathbf{S} - \lambda_{a}\mathbf{T})
\longrightarrow \mathbf{T}^{-1},
\end{eqnarray*}
must hold for the property to hold. We will first embark on
rewriting this implied convergence behavior to a standard form. Note
that we can rewrite such that, equivalently,
\begin{eqnarray*}
&\mathbf{T}^{1/2}\left\{\left[\lambda_{a}\mathbf{I}_{p} +
\frac{1}{4}(\mathbf{S} -
\lambda_{a}\mathbf{T})^{2}\right]^{1/2} +
\frac{1}{2}(\mathbf{S} - \lambda_{a}\mathbf{T})\right\}\mathbf{T}^{1/2} \\
= & \left[\lambda_{a}\mathbf{T}^{2} +
\frac{1}{4}(\tilde{\mathbf{S}} -
\lambda_{a}\mathbf{T}^{2})^{2}\right]^{1/2} +
\frac{1}{2}(\tilde{\mathbf{S}} -
\lambda_{a}\mathbf{T}^{2}) \longrightarrow \mathbf{I}_{p},
\end{eqnarray*}
where $\tilde{\mathbf{S}} =
\mathbf{T}^{1/2}\mathbf{S}\mathbf{T}^{1/2}$, must hold for the
property to hold. Note that the term
$[\lambda_{a}\mathbf{T}^{2} +
\frac{1}{4}(\tilde{\mathbf{S}} -
\lambda_{a}\mathbf{T}^{2})^{2}]^{1/2} +
\frac{1}{2}(\tilde{\mathbf{S}} -
\lambda_{a}\mathbf{T}^{2})$ can be rewritten as:
\begin{eqnarray*}
\left[\frac{1}{4}\left( \lambda_{a}\mathbf{T}^{2} +
2\mathbf{I}_{p} - \tilde{\mathbf{S}} \right)^{2} + \left(
\tilde{\mathbf{S}} - \mathbf{I}_{p} \right)\right]^{1/2} -
\frac{1}{2}\left( \lambda_{a}\mathbf{T}^{2} +
2\mathbf{I}_{p} - \tilde{\mathbf{S}} \right) + \mathbf{I}_{p},
\end{eqnarray*}
implying that the problem can be reduced to proving
\begin{eqnarray}\label{Limitt}
\lim_{\lambda_{a}\rightarrow\infty^{-}} \left[
\mathbf{B}^{2}(\lambda_{a}) + \left( \tilde{\mathbf{S}} -
\mathbf{I}_{p} \right) \right]^{1/2}  -
\mathbf{B}(\lambda_{a}) = \boldsymbol{0},
\end{eqnarray}
where $\mathbf{B}(\lambda_{a}) = \frac{1}{2}\left(
\lambda_{a}\mathbf{T}^{2} + 2\mathbf{I}_{p} -
\tilde{\mathbf{S}} \right)$.

To prove this invoke Weyl's eigenvalue inequality \cite{Weyl12}. Let $\mathbf{A}$, $\mathbf{B}$, $\mathbf{C} = \mathbf{A} + \mathbf{B}$ be real, symmetric $p \times p$ matrices with eigenvalues
$\alpha_1 \geq \alpha_2 \geq \ldots \geq \alpha_p$, $\beta_1 \geq \beta_2 \geq \ldots \geq \beta_p$, and $\gamma_1 \geq \gamma_2 \geq \ldots \geq \gamma_p$, respectively. Weyl's result then states:
\begin{eqnarray*}
\alpha_j + \beta_p \leq \gamma_j \leq \alpha_j + \beta_1 \qquad \mbox{ for all } j.
\end{eqnarray*}
Applying this inequality to $\mathbf{C}(\lambda) = \lambda \mathbf{A} + \mathbf{B}$ with $\lambda > 0$ (where $\lambda$ is used generically) and $\mathbf{A}$ and $\mathbf{B}$ as before, we obtain:
\begin{eqnarray*}
\lambda \alpha_j + \beta_p \leq \gamma_j (\lambda) \leq \lambda \alpha_j + \beta_1 \qquad \mbox{ for all } j.
\end{eqnarray*}
Divide by $\lambda$ and let $\lambda$ tend to infinity (from the left), which is immediately seen to imply:
\begin{eqnarray*}
\lim_{\lambda \rightarrow \infty^{-}} \frac{1}{\lambda} \gamma_j (\lambda) & = &  \alpha_j \qquad \mbox{ for all } j.
\end{eqnarray*}
Put differently, the eigenvalues of $\mathbf{C}(\lambda)$ tend to those of $\lambda \mathbf{A}$. Application of Weyl's eigenvalue inequality and the consequence derived above thus warrant
that
\begin{eqnarray}\nonumber
\lim_{\lambda_{a}\rightarrow\infty^{-}} \left[
\mathbf{B}^{2}(\lambda_{a}) + \left( \tilde{\mathbf{S}} -
\mathbf{I}_{p} \right) \right]^{1/2}  -
\mathbf{B}(\lambda_{a}) = \boldsymbol{0},
\end{eqnarray}
as indeed needed to be proven. \QEDE
\end{poP1}

\newproof{poC1}{Proof of Corollary \ref{RidgeAltIIColl}}
\begin{poC1}\label{RidgeAltIICollProof}
We first need to show that (\ref{RidgeAltII}) stems properly as the unique maximizer of the log-likelihood (\ref{form.loglik}) amended with (\ref{RidgePenal}) under $\mathbf{T} = \boldsymbol{0}$. We thus define the penalized log-likelihood:
\begin{equation}\nonumber
\mathcal{L}^{p}(\mathbf{\Omega};\mathbf{S}, \lambda_{a}) \propto \ln|\mathbf{\Omega}| - \mbox{tr}(\mathbf{S \Omega}) - \frac{\lambda_{a}}{2}\mbox{tr}(\mathbf{\Omega}^{\mathrm{T}}\mathbf{\Omega}).
\end{equation}
Taking the derivative of $\mathcal{L}^{p}(\mathbf{\Omega};\mathbf{S}, \lambda_{a})$ w.r.t. $\mathbf{\Omega}$ gives:
\begin{equation}\nonumber
\frac{\partial \, \mathcal{L}^{p}(\mathbf{\Omega};\mathbf{S}, \lambda_{a})}{\partial \, \mathbf{\Omega}} =
\left(\mathbf{\Omega}^{-1} - \mathbf{S} - \lambda_{a}\mathbf{\Omega}\right) \circ
\left( 2\mathbf{J}_{p} -  \mathbf{I}_{p} \right),
\end{equation}
which is $\boldsymbol{0}$ only when
\begin{equation}\nonumber
\mathbf{\Omega}^{-1} - \mathbf{S} - \lambda_{a}\mathbf{\Omega} = \boldsymbol{0}.
\end{equation}
A strategy analogous to the one used in the proof of Lemma \ref{GenRidgeAltLemma} will give the desired expression (\ref{RidgeAltII}). With regard to the properties of this estimator:

(i) Let $d(\cdot)_{jj}$ denote the $j$'th eigenvalue of the matrix term in brackets $(\cdot)$. Notice
\begin{eqnarray*}
d\left\{[\hat{\mathbf{\Omega}}^{\mathrm{II}a}(\lambda_{a})]^{-1}\right\}_{jj} =
d\left(\frac{1}{2}\mathbf{S}\right)_{jj}
+
\sqrt{\left[d\left(\frac{1}{2}\mathbf{S}\right)_{jj}\right]^2
+ \lambda_{a}}
\, \, \, > \, \, \, 0,
\end{eqnarray*}
when $\lambda_{a} > 0$, implying $\hat{\mathbf{\Omega}}^{\mathrm{II}a}(\lambda_{a})$ is p.d. for any $\lambda_{a} \in (0,\infty)$.

(ii) The right-hand limit is immediate as:
\begin{eqnarray*}
\hat{\mathbf{\Omega}}^{\mathrm{II}a}(0) = \left\{\left[0\mathbf{I}_{p} + \frac{1}{4}\mathbf{S}^{2}\right]^{1/2} + \frac{1}{2}\mathbf{S}\right\}^{-1} = \mathbf{S}^{-1}.
\end{eqnarray*}

(iii) For the left-hand limit we note that as $\lambda_{a}$ approaches $\infty$,
\begin{eqnarray*}
[\hat{\mathbf{\Omega}}^{\mathrm{II}a}(\lambda_{a})]^{-1} = \left[\lambda_{a}\mathbf{I}_{p} + \frac{1}{4}\mathbf{S}^{2}\right]^{1/2} + \frac{1}{2}\mathbf{S}
\end{eqnarray*}
becomes a diagonally dominant matrix with near infinite diagonal
values. The inverse of which must necessarily approach the
null-matrix. \QEDE
\end{poC1}

\newproof{poP2}{Proof of Proposition \ref{prop.consistency}}
\begin{poP2}\label{prop.consistencyProof}
First, note:
\begin{align*}
\mathbb{E} \left( \|  \hat{\mathbf{\Sigma}}_n^{\mathrm{I}a} (\lambda_{a,n}) - \mathbf{\Sigma} \|_F^2 \right)  = & ~\mathbb{E} \left\{ \mbox{tr}\left[ \left( \hat{\mathbf{\Sigma}}_n^{\mathrm{I}a} (\lambda_{a,n}) - \mathbf{\Sigma} \right)^{\mathrm{T}} \left( \hat{\mathbf{\Sigma}}_n^{\mathrm{I}a} (\lambda_{a,n}) - \mathbf{\Sigma} \right)\right] \right\}
\\
 = & ~\mbox{tr} \left\{ \mbox{Var}  \left[  \hat{\mathbf{\Sigma}}_n^{\mathrm{I}a} (\lambda_{a,n}) \right] \right\}
\\
& ~ - \mbox{tr} \left\{  \mathbf{\Sigma} \, \mathbb{E} \left[  \hat{\mathbf{\Sigma}}_n^{\mathrm{I}a} (\lambda_{a,n}) - \mathbf{\Sigma} \right] \right\}
  \mbox{tr} \left\{ \mathbb{E} \left[   \hat{\mathbf{\Sigma}}_n^{\mathrm{I}a} (\lambda_{a,n}) - \mathbf{\Sigma} \right]   \mathbf{\Sigma}  \right\}.
\end{align*}
By virtue of Lemma \ref{prop.asympBias} and the continuity of the trace the latter two terms vanish as $n \rightarrow \infty^{-}$. It remains to be shown that the first term converges to zero. To this end note that the almost sure convergence of $\lambda_{a,n}$ to zero implies $\lim_{n \rightarrow \infty^{-}} P [\lambda_{a,n} d(\mathbf{T})_{11}  < d(\mathbf{S}_n)_{pp}] = 1$ and $\mathbf{S}_n - \lambda_{a,n} \mathbf{T} \succcurlyeq \mathbf{0}$ with probability 1 as $n$ tends to infinity. Thus, in the limit:
\begin{eqnarray*}
\hat{\mathbf{\Sigma}}_n^{\mathrm{I}a} (\lambda_{a,n}) & = & \left[\lambda_{a,n} \mathbf{I}_{p} + \frac{1}{4} (\mathbf{S}_n - \lambda_{a,n} \mathbf{T})^2\right]^{1/2} + \frac{1}{2} (\mathbf{S}_n - \lambda_{a,n} \mathbf{T})
\\
& \preccurlyeq & \left\{\left[\sqrt{\lambda_{a,n}} \mathbf{I}_{p} + \frac{1}{2} (\mathbf{S}_n - \lambda_{a,n} \mathbf{T})\right]^2\right\}^{1/2} +
\frac{1}{2} (\mathbf{S}_n - \lambda_{a,n} \mathbf{T})
\\
& = & \mathbf{S}_n  + \sqrt{\lambda_{a,n}} \mathbf{I}_{p}  - \lambda_{a,n} \mathbf{T}.
\end{eqnarray*}
From this it follows that
\begin{eqnarray*}
\mbox{tr} \left[ \left(\mathbf{S}_n  + \sqrt{\lambda_{a,n}} \mathbf{I}_{p}  - \lambda_{a,n} \mathbf{T}\right)^2 \right] \succcurlyeq  \mbox{tr} \left\{ \left[ \hat{\mathbf{\Sigma}}^{\mathrm{I}a}_n (\lambda_{a,n}) \right]^2\right\}
\end{eqnarray*}
as $n\rightarrow \infty^{-}$, which in turn gives:
\begin{eqnarray*}
\mbox{tr} \left[ \mbox{Var}\left( \mathbf{S}_n  + \sqrt{\lambda_{a,n}} \mathbf{I}_{p}  - \lambda_{a,n} \mathbf{T} \right) \right] \succcurlyeq  \mbox{tr} \left\{ \mbox{Var}\left[  \hat{\mathbf{\Sigma}}_n^{\mathrm{I}a} (\lambda_{a,n}) \right]  \right\} \quad \mbox{as } n\rightarrow \infty^{-}.
\end{eqnarray*}
The assumptions on $\lambda_{a,n}$, $\mathbf{S}_n$ and their covariance imply that the left-hand side tends to zero. Finally, the dominated convergence theorem warrants that the right-hand side too converges to zero as $n \rightarrow \infty^{-}$. \QEDE
\end{poP2}

The proof of Proposition \ref{AltIEIGprop} will be based on the target $\mathbf{T} = \mathbf{I}_{p}$.
The extension to a general p.d. target scalar matrix is
straightforward as it is a direct consequence, but notationally slightly more cumbersome.

\newproof{poP3}{Proof of Proposition \ref{AltIEIGprop}}
\begin{poP3}\label{AltIEIGpropProof}
Note that $\hat{\mathbf{\Omega}}^{\mathrm{I}}(\lambda_{\mathrm{I}})$ can be decomposed as:
\begin{equation*}
[ \hat{\mathbf{\Omega}}^{\mathrm{I}}(\lambda_{\mathrm{I}}) ]^{-1}
 =  \mathbf{V} [(1 - \lambda_{\mathrm{I}}) \mathbf{D} +  \lambda_{\mathrm{I}}  \mathbf{I}_{p}] \mathbf{V}^\mathrm{T}.
\end{equation*}
Juxtaposing this expression with (\ref{TypeIDecomp}) while writing $d_{jj} = (\mathbf{D})_{jj}$, we are after establishing
\begin{eqnarray*}
\sqrt{ \lambda_{a} + \frac{1}{4} \left( d_{jj} - \lambda_{a} \right)^2} +
\frac{1}{2} \left( d_{jj} - \lambda_{a} \right)  &\stackrel{?}{> = <}&  \frac{1}{1+\lambda_{a}} d_{jj} +
\frac{\lambda_{a}}{1+\lambda_{a}},
\end{eqnarray*}
which after some ready algebra can be rewritten as:
\begin{eqnarray*}
\sqrt{  \varphi_{jj}(\lambda_{a})^2 +
d_{jj} -1 } & \stackrel{?}{> = < } & \frac{1}{1+\lambda_{a}} (d_{jj} -1)
+  \varphi_{jj}(\lambda_{a}),
\end{eqnarray*}
with $\varphi_{jj}(\lambda_{a}) = \frac{1}{2} \lambda_{a} - \frac{1}{2} d_{jj} + 1$.
Squaring both sides and simplifying the problem becomes:
\begin{eqnarray*}
d_{jj} -1 & \stackrel{?}{> = < } & \frac{1}{(1+\lambda_{a})^2} (d_{jj} -1)^2 + d_{jj} -1  -
\frac{1}{1+\lambda_{a}} ( d_{jj} - 1)^2,
\end{eqnarray*}
which reduces to establishing the sign of:
\begin{eqnarray*}
\frac{(d_{jj} -1)^2}{(1+\lambda_{a})^2} -
\frac{( d_{jj} - 1)^2}{1+\lambda_{a}}.
\end{eqnarray*}
The solution to which is readily found to be:
\begin{eqnarray*}
0  \geq  \frac{(d_{jj} -1)^2}{(1+\lambda_{a})^2} -
\frac{( d_{jj} - 1)^2}{1+\lambda_{a}}
 =  - \frac{\lambda_{a} \, (d_{jj} -1)^2 }{(1+\lambda_{a})}.
\end{eqnarray*}
Consequently, the alternative estimator
$\hat{\mathbf{\Omega}}^{\mathrm{I}a}(\lambda_{a})$ displays
shrinkage of the eigenvalues of $\mathbf{S}^{-1}$ that is at least
as heavy as the shrinkage propagated by the archetypal estimator
$\hat{\mathbf{\Omega}}^{\mathrm{I}}(\lambda_{\mathrm{I}})$ . \QEDE
\end{poP3}

\newproof{poP4}{Proof of Proposition \ref{AltIIEIGprop}}
\begin{poP4}\label{AltIIEIGpropProof}
Note that the decomposition of the original ridge estimator of the
second type is
\begin{equation*}
[ \hat{\mathbf{\Omega}}^{\mathrm{II}}(\lambda_{\mathrm{II}}) ]^{-1}
 = \mathbf{V} ( \lambda_{\mathrm{II}}  \mathbf{I}_{p}   +
\mathbf{D} ) \mathbf{V}^\mathrm{T}.
\end{equation*}
Then, when writing $d_{jj} = (\mathbf{D})_{jj}$ while juxtaposing
the above expression with (\ref{TypeIIDecomp}), we have:
\begin{eqnarray*}
\lambda_{\mathrm{II}} + d_{jj} & \geq &
\sqrt{\lambda_{\mathrm{II}}^2 + \frac{1}{4} d_{jj}^2} + \frac{1}{2}
d_{jj},
\end{eqnarray*}
as follows directly from $(\lambda_{\mathrm{II}} + \frac{1}{2}
d_{jj})^2  \geq \lambda_{\mathrm{II}}^2 + \frac{1}{4} d_{jj}^2$.
This indicates the archetypal estimator
$\hat{\mathbf{\Omega}}^{\mathrm{II}}(\lambda_{\mathrm{II}})$ displaying
shrinkage of the eigenvalues of $\mathbf{S}^{-1}$ that is at least
as heavy as the shrinkage propagated by the alternative estimator
$\hat{\mathbf{\Omega}}^{\mathrm{II}a}(\lambda_{a})$. \QEDE
\end{poP4}

\newproof{poC2}{Proof of Corollary \ref{AltIILIKEcor}}
\begin{poC2}\label{AltIILIKEcorProof}
Note:
\begin{eqnarray*}
\mathcal{L}[\hat{\mathbf{\Omega}}^{\mathrm{II}}
(\lambda_{\mathrm{II}}); \mathbf{S}] \propto
\ln|\hat{\mathbf{\Omega}}^{\mathrm{II}} (\lambda_{\mathrm{II}}) | -
\mbox{tr} [ \mathbf{S} \hat{\mathbf{\Omega}}^{\mathrm{II}}
(\lambda_{\mathrm{II}}) ]
\propto  - \sum_{j=1}^p \ln(\lambda_{\mathrm{II}} + d_{jj}) -
\sum_{j=1}^p \frac{d_{jj}}{\lambda_{\mathrm{II}} + d_{jj}}.
\end{eqnarray*}
Similarly:
\begin{eqnarray*}
\mathcal{L}[\hat{\mathbf{\Omega}}^{\mathrm{II}a}
(\lambda_{a}); \mathbf{S}] \propto - \sum_{j=1}^p \ln
[\gamma_{jj}(\lambda_{\mathrm{II}})] - \sum_{j=1}^p
\frac{d_{jj}}{\gamma_{jj}(\lambda_{\mathrm{II}})},
\end{eqnarray*}
where $\gamma_{jj}(\lambda_{\mathrm{II}}) =
\sqrt{\lambda_{\mathrm{II}}^2 + \frac{1}{4}d_{jj}^2} + \frac{1}{2}
d_{jj}$. It then suffices to show that
\begin{eqnarray*}
\ln(\lambda_{\mathrm{II}} + d_{jj}) -
\ln[\gamma_{jj}(\lambda_{\mathrm{II}})] +
\frac{d_{jj}}{\lambda_{\mathrm{II}} + d_{jj}}  -
\frac{d_{jj}}{\gamma_{jj}(\lambda_{\mathrm{II}})} & \geq & 0.
\end{eqnarray*}
Using $\ln(1+x) \geq x / (1+x)$ and $d_{jj} + \lambda_{\mathrm{II}}
\geq \gamma_{jj}(\lambda_{\mathrm{II}}) \geq d_{jj}$ (Proposition \ref{AltIIEIGprop}), the
manipulations below prove this:
\begin{eqnarray*}
\ln \left(\frac{\lambda_{\mathrm{II}} +
d_{jj}}{\gamma_{jj}(\lambda_{\mathrm{II}})} \right) +
\frac{d_{jj}}{\lambda_{\mathrm{II}} + d_{jj}}  -
\frac{d_{jj}}{\gamma_{jj}(\lambda_{\mathrm{II}})} & \geq &
\\
\frac{\lambda_{\mathrm{II}} + d_{jj} -
\gamma_{jj}(\lambda_{\mathrm{II}})}{\lambda_{\mathrm{II}} + d_{jj} }
+ \frac{d_{jj}}{\lambda_{\mathrm{II}} + d_{jj}}  -
\frac{d_{jj}}{\gamma_{jj}(\lambda_{\mathrm{II}})} & = &
\\
\frac{ [\gamma_{jj}(\lambda_{\mathrm{II}}) - d_{jj}] [d_{jj} +
\lambda_{\mathrm{II}} - \gamma_{jj}(\lambda_{\mathrm{II}})]  }{
\gamma_{jj}(\lambda_{\mathrm{II}}) (\lambda_{\mathrm{II}} + d_{jj})
} & \geq & 0.
\end{eqnarray*} \QEDE
\end{poC2}

%---- References
%-----------------------------------------------------------------------------------
\section*{References}
\bibliographystyle{elsarticle-num}
\bibliography{vanWieringenPeeters2015_Ridge}

\begin{thebibliography}{10}
\expandafter\ifx\csname url\endcsname\relax
  \def\url#1{\texttt{#1}}\fi
\expandafter\ifx\csname urlprefix\endcsname\relax\def\urlprefix{URL }\fi
\expandafter\ifx\csname href\endcsname\relax
  \def\href#1#2{#2} \def\path#1{#1}\fi

\bibitem{YL07}
M.~Yuan, Y.~Lin, Model selection and estimation in the {Gaussian} graphical
  model, Biometrika 94 (2007) 19--35.

\bibitem{Bane2008}
O.~Banerjee, L.~{El Ghaoui}, A.~{d'Aspremont}, Model selection through sparse
  maximum likelihood estimation for multivariate {G}aussian or binary data,
  Journal of Machine Learning Research 9 (2008) 485--516.

\bibitem{Frie2008}
J.~Friedman, T.~Hastie, R.~Tibshirani, Sparse inverse covariance estimation
  with the graphical lasso, Biostatistics 9 (2008) 432--441.

\bibitem{Yuan08}
M.~Yuan, Efficient computation of $\ell_{1}$ regularized estimates in
  {Gaussian} graphical models, Journal of Computational and Graphical
  Statistics 17 (2008) 809--826.

\bibitem{Fu1998}
W.~J. Fu, Penalized regressions: {T}he bridge versus the lasso, Journal of
  Computational and Graphical Statistics 7 (1998) 397--416.

\bibitem{Ledo2004}
O.~Ledoit, M.~Wolf, A well-conditioned estimator for large-dimensional
  covariance matrices, Journal of Multivariate Analysis 88 (2004) 365--411.

\bibitem{SS05}
J.~Sch\"{a}fer, K.~Strimmer, A shrinkage approach to large-scale covariance
  matrix estimation and implications for functional genomics, Statistical
  Applications in Genetics and Molecular Biology 4 (2005) art. 32.

\bibitem{Hoer1970}
A.~E. Hoerl, R.~Kennard, Ridge regression: {B}iased estimation for
  nonorthogonal problems, Technometrics 12 (1970) 55--67.

\bibitem{Wart08}
D.~Warton, Penalized normal likelihood and ridge regularization of correlation
  and covariance matrices, Journal of the American Statistical Association 103
  (2008) 340--349.

\bibitem{Wit09}
D.~M. Witten, R.~Tibshirani, Covariance-regularized regression and
  classification for high-dimensional problems, Journal of the Royal
  Statistical Society, Series B 71 (2009) 615--636.

\bibitem{Wish}
G.~Letac, H.~Massam, All invariant moments of the {Wishart} distribution,
  Scandinavian Journal of Statistics 31 (2004) 295--318.

\bibitem{Vaart98}
A.~W. {van der Vaart}, {Asymptotic Statistics}, {Cambridge University Press,
  Cambridge}, 1998.

\bibitem{Abbruz2014}
A.~Abbruzzo, I.~Vuja\v{c}i\'{c}, E.~Wit, A.~M. Mineo, Generalized information
  criterion for model selection in penalized graphical models,
  \texttt{arXiv:1403.1249v1 [stat.ME]} (2014).

\bibitem{Aik73}
H.~Akaike, Information theory and an extension of the maximum likelihood
  principle, in: B.~N. Petrov, F.~Csaki (Eds.), Second International Symposium
  on Information Theory, Akademiai Kaido, Budapest, 1973, pp. 267--281.

\bibitem{Lian2011}
H.~Lian, Shrinkage tuning parameter selection in precision matrices estimation,
  Journal of Statistical Planning and Inference 141 (2011) 2839--2848.

\bibitem{Vuja2014}
I.~Vuja\v{c}i\'{c}, A.~Abbruzzo, E.~C. Wit, A computationally fast alternative
  to cross-validation in penalized {Gaussian} graphical models,
  \texttt{arXiv:1309.621v2 [stat.ME]} (2014).

\bibitem{Ledo2003}
O.~Ledoit, M.~Wolf, Improved estimation of the covariance matrix of stock
  returns with an application to portfolio selection, Journal of Empirical
  Finance 10 (2003) 603--621.

\bibitem{DK01}
M.~J. Daniels, R.~E. Kass, Shrinkage estimators for covariance matrices,
  Biometrics 57 (2001) 1173--1184.

\bibitem{glassoMAN}
J.~Friedman, T.~Hastie, R.~Tibshirani,
  \href{http://CRAN.R-project.org/package=glasso}{\texttt{glasso}: Graphical
  lasso-estimation of Gaussian graphical models}, \texttt{R} package, version
  1.7 (2011).
\newline\urlprefix\url{http://CRAN.R-project.org/package=glasso}

\bibitem{whittaker}
J.~Whittaker, {Graphical Models in Applied Multivariate Statistics}, {John
  Wiley \& Sons Ltd., Chichester}, 1990.

\bibitem{EfronLocFDR}
B.~Efron, R.~Tibshirani, J.~D. Storey, V.~Tusher, Empirical {B}ayes analysis of
  a microarray experiment, Journal of the American Statistical Association 96
  (2001) 1151--1160.

\bibitem{Hotel}
H.~Hotelling, New light on the correlation coefficient and its transforms,
  Journal of the Royal Statistical Society, Series B 15 (1953) 193--232.

\bibitem{SS052}
J.~Sch\"{a}fer, K.~Strimmer, An empirical {Bayes} approach to inferring
  large-scale gene association networks, Bioinformatics 21 (2005) 754--764.

\bibitem{EfronLSbook}
B.~Efron, {Large-Scale Inference: Empirical Bayes Methods for Estimation,
  Testing, and Prediction}, {Cambridge University Press, Cambridge}, 2010.

\bibitem{LSadjust}
S.~Boyd, L.~Xiao, Least-squares covariance matrix adjustment, SIAM Journal on
  Matrix Analysis and Applications 27 (2005) 532--546.

\bibitem{BioCon}
R.~C. Gentleman, V.~J. Carey, D.~M. Bates, B.~M. Bolstad, M.~Dettling,
  S.~Dudoit, B.~Ellis, L.~Gautier, Y.~Ge, J.~Gentry, K.~Hornik, T.~Hothorn,
  W.~Huber, S.~Iacus, R.~Irizarry, F.~Leisch, L.~Cheng, M.~Maechler, A.~J.
  Rossini, G.~Sawitizki, C.~Smith, G.~Smyth, L.~Tierney, J.~Y.~H. Yang,
  J.~Zhang, Bioconductor: Open software development for computational biology
  and bioinformatics, Genome Biology 5 (2004) R80.

\bibitem{BreastData}
M.~Schr\"{o}der, B.~Haibe-Kains, A.~Culhane, C.~Sotiriou, G.~Bontempi,
  J.~Quackenbush, \href{http://compbio.dfci.harvard.edu/}{breastCancerMAINZ;
  breastCancerTRANSBIG; breastCancerUNT; breastCancerUPP; breastCancerVDX},
  \texttt{R} packages, versions 1.0.6 (2011).
\newline\urlprefix\url{http://compbio.dfci.harvard.edu/}

\bibitem{QuackB12}
B.~{Haibe-Kains}, C.~Desmedt, S.~Loi, A.~C. Culhane, G.~Bontempi,
  J.~Quackenbush, C.~Sotiriou, A three-gene model to robustly identify breast
  cancer molecular subtypes, Journal of the National Cancer Institute 104
  (2012) 311--325.

\bibitem{KEGG}
M.~Kanehisa, S.~Goto, {KEGG: Kyoto Encyclopedia of Genes and Genomes}, Nucleic
  Acids Research 28 (2000) 27--30.

\bibitem{MBC}
L.~Pecorino, {Molecular Biology of Cancer: Mechanisms, Targets and
  Therapeutics}, 3rd Edition, {Oxford University Press, Oxford}, 2012.

\bibitem{Vogelp53}
B.~Vogelstein, S.~Sur, C.~Prives, {p53: The} most frequently altered gene in
  human cancers, Nature Education 3 (2010) 6.

\bibitem{Hallmarks}
D.~Hanahan, R.~A. Weinberg, The hallmarks of cancer, Cell 100 (2000) 57--70.

\bibitem{Jor81}
K.~G. J\"{o}reskog, Analysis of covariance structures, Scandinavian Journal of
  Statistics 8 (1981) 65--92.

\bibitem{Won2013}
J.~H. Won, J.~Lim, S.~J. Kim, B.~Rajaratnam, Condition-number-regularized
  covariance estimation, Journal of the Royal Statistical Society, Series B 75
  (2013) 427--450.

\bibitem{rags}
C.~F.~W. Peeters, W.~N. {van Wieringen},
  \href{http://cran.r-project.org/web/packages/rags2ridges/index.html}{\texttt{rags2ridges}:
  Ridge estimation of precision matrices from high-dimensional data},
  \texttt{R} package, Version 1.3 (2014).
\newline\urlprefix\url{http://cran.r-project.org/web/packages/rags2ridges/index.html}

\bibitem{Rman}
{\texttt{R} Development Core Team},
  \href{http://www.R-project.org/}{\texttt{R}: A Language and Environment for
  Statistical Computing}, \texttt{R} Foundation for Statistical Computing,
  Vienna, Austria (2011).
\newline\urlprefix\url{http://www.R-project.org/}

\bibitem{Serre02}
D.~Serre, {Matrices: Theory and Applications}, Springer, New York, 2002.

\bibitem{Weyl12}
H.~Weyl, Das asymptotische {Verteilungsgesetz} der {Eigenwerte} linearer
  partieller {Differentialgleichungen} (mit einer {Anwendung} auf die {Theorie}
  der {Hohlraumstrahlung}), Mathematische Annalen 71 (1912) 441--479.

\end{thebibliography}

\end{document}